\documentclass[twocolumn,floatfix,pre,aps,showpacs]{revtex4}
\usepackage{graphicx,amsmath,amssymb}

\begin{document}

\title{Path-integral Monte Carlo simulation of time-reversal noninvariant bulk systems
with a case study of rotating Yukawa gases}

\author{Tam\'as Haidekker Galambos$^{1,2,3}$ and Csaba T\H oke$^{1,2}$}
\affiliation{$^{1}$BME-MTA Exotic Quantum Phases ``Lend\"ulet" Research Group, Budapest University of Technology and Economics, Institute of Physics, Budafoki \'ut 8, H-1111 Budapest, Hungary}
\affiliation{$^{2}$Department of Theoretical Physics,Budapest University of Technology and Economics, Institute of Physics, Budafoki \'ut 8, H-1111 Budapest, Hungary}
\affiliation{$^{3}$Department of Physics, University of Basel, Klingelbergstrasse 82, CH-4056 Basel, Switzerland}

\date{\today}

\begin{abstract}
We elaborate on the methodology to simulate bulk systems in the absence of time-reversal symmetry by the
phase-fixed path-integral Monte Carlo method under (possibly twisted) periodic boundary conditions.
Such systems include two-dimensional electrons in the quantum Hall regime
and rotating ultracold Bose and Fermi gases; time-reversal symmetry is broken by an external magnetic field
and the Coriolis force, respectively.
We provide closed-form expressions in terms of Jacobi elliptic functions for the thermal density matrix
(or the Euclidean propagator) of a single particle on a flat torus under very general conditions.
We then modify the multi-slice sampling method in order to sample paths by
the magnitude of the complex-valued thermal density matrix.
Finally, we demonstrate that these inventions let us study the vortex melting process of a two-dimensional
Yukawa gas in terms of the de Boer interaction strength parameter, temperature, and rotation (Coriolis force).
The bosonic case is relevant to ultracold Fermi-Fermi mixtures of widely different masses under rotation.
\end{abstract}

\maketitle

\section{Introduction}

The path-integral Monte Carlo (PIMC) method \cite{Ceperley95} lets us simulate many-body systems at finite temperature
in a controlled manner.
Equilibrium properties are obtained from the many-body density matrix
\begin{equation} 
\label{eq:dm}
\rho(R,R';\beta) = \sum_{n}e^{-\beta\epsilon_n}\Psi_n(R)\Psi^\ast_n(R'),
\end{equation} 
where $R\equiv(\mathbf r_1,\mathbf r_2,\dots,\mathbf r_N)$ collects $dN$ particle coordinates,
$d$ is the dimensionality of the system, $N$ is the number of particles,
$\{\Psi_n\}$ is a complete set of many-body eigenstates, and $\{\epsilon_n\}$ are the corresponding energies.
The convolution identity of the density matrix,
\begin{equation}
\label{eq:conv}
\rho(R,R';\beta_1+\beta_2) = \int dR'' \rho(R,R'';\beta_1)\rho(R'',R';\beta_2),
\end{equation}
is applied iteratively to yield the imaginary-time path-integral representation
\begin{multline}
\label{eq:pathint}
\rho(R,R';\beta) = \int dR_1\cdots\int dR_{M-1} \rho(R,R_1;\tau)\\
\times\rho(R_1,R_2;\tau)\dots\rho(R_{M-1},R';\tau).
\end{multline}
Here, the time-step $\tau\equiv\beta/M$ corresponds to a much higher temperature than the system temperature.
The high-temperature density matrix that connects adjacent slices,
$\rho(R_{m-1},R_m;\tau)$, can be approximated by several plausible schemes \cite{Ceperley95}.
Estimators of physical quantities are defined by integrals that involve $\rho(R,R';\beta)$;
in most cases the diagonal element $\rho(R,R;\beta)$ is sufficient.
The Metropolis-Hastings Monte Carlo method \cite{Metropolis53,Hastings70} is applicable to path integration
if the product of high-temperature density matrices in Eq.~(\ref{eq:pathint})
can be interpreted as a probability density function.

For time-reversal invariant bosonic systems this always holds,
and PIMC is an unbiased and essentially exact method in this case.
For fermions, however, the notorious sign problem arises, because the contribution of a particular path
can have either sign due to the presence of nondiagonal factors $\rho(R_{m-1},R_m;\tau)$
in the integrand of estimators.
The generic means to overcome this problem,
the use of restricted or constrained paths that avoid the nodal surfaces of a preconceived
trial many-body density matrix \cite{Ceperley91,Ceperley96}, makes PIMC variational in character.

On the other hand, if time-reversal is not a symmetry of the system,
either because charged particles are exposed to an external magnetic field or the system is rotated,
the density matrices are complex-valued in general
and hence the prescription of the nodal surfaces is insufficient.
A consequent method would be to sample paths by the probability density function (PDF)
$\prod_{m=1}^{M}|\rho(R_{m-1},R_m;\tau)|$ (here we assume integration with the diagonal
density matrix as the kernel of the estimator, and we define $R=R'\equiv R_0=R_M$),
and sum them up with the complex phase factor $\prod_{m=1}^{M}\frac{\rho(R_{m-1},R_m;\tau)}{|\rho(R_{m-1},R_m;\tau)|}$.
This procedure would result in a more severe form of the sign problem: contributions with different
phase factors would cancel almost completely.
This issue is equally severe for bosons and fermions, and it arises even
in the nonphysical case of distinguishable particles (``bolzmannons'').
In analogy to the phase-fixing extension \cite{Ortiz93,Bolton96} of zero-temperature
methods such as diffusion quantum Monte Carlo \cite{Foulkes01},
phase fixing is an obvious route to adapt PIMC to such problems.
Unlike the case of zero-temperature methods, the function whose phase needs to be fixed
is the many-body density matrix in Eq.~(\ref{eq:dm}), not a wave function.
While the fixed-phase extension of the PIMC method is often mentioned in the literature \cite{Akkineni08},
it is hardly ever applied, in contrast to the similar extension of zero-temperature methods
\cite{Ortiz93,Melik-Alaverdian95,Bolton96,Melik-Alaverdian97}.

We address several issues related to the use of PIMC in time-reversal non-invariant bulk systems.
(Finite systems such as quantum dots are not our primary interest here.)
First, if we want to simulate bulk systems consequently, we have to use periodic boundary conditions,
possibly with twist angles that let us reduce finite-size effects such as shell effects
in finite-size representations of Fermi liquids \cite{Lin01}, which have analogs in strongly
correlated electron systems in magnetic fields \cite{Shao15}.
One should base any PIMC simulation on the single-particle thermal density matrix (equivalently, kinetic action)
that is exact under the chosen boundary conditions.
We show that the free propagation of a charged particle (equivalently, the thermal density matrix)
on a flat torus subjected to a perpendicular magnetic field already exhibits a rather rich structure,
although these patterns lose their significance for small imaginary times or large system sizes.
This result lets us define the kinetic action in a way that is compatible with the torus.

The PIMC method is applicable beyond toy models only because the sampling of paths could be made efficient
by the introduction of multi-slice moves.
These replace entire segments of the path \cite{Pollock84}
according to the PDF $\prod_{m=1}^{M}\rho(R_{m-1},R_m;\tau)$.
If, however, the density matrix is complex-valued and the probability density of paths
is determined by its magnitude, the familiar bisection method \cite{Ceperley95}
that relies on the L\'evy construction of a Brownian bridge, runs into difficulties
because the convolution property in Eq.~(\ref{eq:conv}) is not applicable to magnitudes.
We elaborate on a modification of the multi-slice move algorithm that takes
the external magnetic field and the periodicity of the torus into account.

Finally, we demonstrate the use of phase-fixed PIMC for bulk systems
in a case study of rotating two-dimensional Yukawa gases.
Yukawa bosons arise either in type-II superconductors, where the Abrikosov vortex
lines interact by a repulsive modified-Bessel-function potential $\propto K_0(r)$ \cite{Nelson89,Magro93,Nordborg97},
or in strongly interacting Fermi-Fermi mixtures of ultracold atoms, if the mass ratio of the
two species, $M/m$, is very far from unity and the motion of both species is confined to two dimensions \cite{Petrov07}.
A flux density can be introduced to cold atomic systems by rotating the gas, a technique that has been applied frequently
in the past two decades \cite{Madison00,AboShaeer01,Hodby01,Haljan01}.
In the model we consider particles that interact via a modified-Bessel-function potential $\propto K_0(r)$.
This is a good approximation also to the inter-atomic interaction in a Fermi-Fermi mixture at sufficiently
long range \cite{Petrov07}.
We do not claim, however, to represent either problem faithfully:
we do not include the nonuniversal short-range repulsion between Fermi-Fermi bound states,
and the inclusion of additional flux density would be difficult to justify for Abrikosov vortices.
We have deliberately chosen this system for computational convenience in order to demonstrate the adequacy of our methodology.
On the one hand, $K_0(r)$ is mildly divergent at short range, thus even the simplest approximation
to the high-temperature density matrix, the primitive action, is a reasonable starting point.
On the other, as $K_0(r)$ decays exponentially at large range, the intricacies of Ewald summation can be avoided.

As a first approach, we use the density matrix of the free Bose and Fermi gases to
fix the phase of the many-body density matrix.
We are encouraged in this by the fact that in the case of the node fixing problem,
which arises analogously for time-reversal invariant fermionic systems, significant
progress was possible both for $^3$He \cite{Ceperley92} and the hydrogen plasma \cite{Pierleoni94,Magro96}
using the nodal surfaces of either the noninteracting system or some well-tested variational ground state wave function.
(The two approaches are somewhat complementary.)
Simple as it is, we demonstrate that phase-fixed PIMC captures the crystallization
of rotating Yukawa bosons and fermions as a function of interaction strength, flux density, and temperature.
We emphasize that unlike for the diffusion Monte Carlo or Green's Function Monte Carlo methods,
no trial wave function of the proper symmetry serves as input to such a calculation;
but we do choose the aspect ratio of the unit cell so that
it can accommodate a finite piece of a triangular lattice.

The paper is structured as follows.
In Sec.~\ref{sec:freedm} we present the density matrix for a single particle in a magnetic field
on the torus, with some mathematical details of the derivation delegated to Appendix \ref{app:dm},
and the considerations of its efficient computation to Appendix \ref{app:comput}.
The adaptation of the multi-slice sampling algorithm is discussed in Sec.~\ref{multislice},
with a detour to periodic, but time-reversal-invariant systems.
Sec.~\ref{yukawa} presents a case study, where the phase-fixed path-integral Monte Carlo method
is applied to rotating systems of two-dimensional Yukawa gases under periodic boundary conditions.
In Sec.~\ref{conclusion} we summarize our results and discuss further research directions.
Appendix \ref{app:pfaction} presents the technical details of the phase-fixing methodology for PIMC.

\section{The thermal density matrix}
\label{sec:freedm}

\begin{figure}[htbp]
\begin{center}
\includegraphics[width=.49\columnwidth]{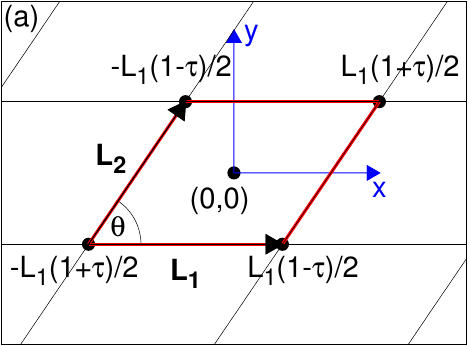}
\includegraphics[width=.49\columnwidth]{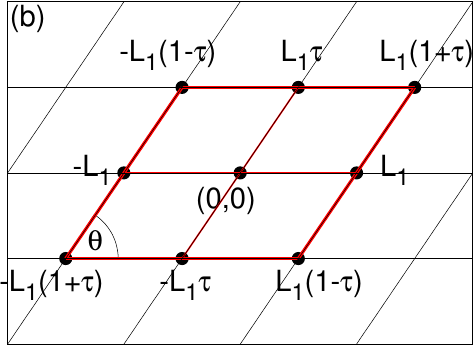}
\end{center}
\caption{\label{fig:domain}
(a) The principal domain of the torus.
We also depict $\mathbf L_1$, $\mathbf L_2$ and $\theta$ as defined in the text;
we identify the plane with the complex plane, and indicate the corners of the
principal domain using the complex parameter $\tau$ defined in Eq.~(\ref{eq:deftau}).
(b) The quadruple domain used for finding the zeros of the density matrix in the low-temperature limit.
}
\end{figure}

We consider a flat torus pierced by a perpendicular magnetic field.
Consider the parallelogram spanned by two nonparallel vectors $\mathbf L_1=(L_1,0)$ and
$\mathbf L_2=(L_2\cos\theta,L_2\sin\theta)$.
A torus is obtained by identifying the opposite sides of this unit cell; cf.\ Fig.~\ref{fig:domain}(a).
We will refer to a similar parallelogram that has the origin as its center as the principal domain.

We use the Landau gauge $\mathbf A = -By\mathbf{\hat x}$ throughout this article.
Electrons are characterized by complex coordinates $z=x+iy$,
and we define
\begin{equation}
\label{eq:deftau}
\tau=\frac{L_2}{L_1}e^{i\theta},
\end{equation}
so that $L_1$ and $L_1\tau$ span the parallelogram on the complex plane.
In the presence of a perpendicular magnetic field, magnetic translations \cite{Zak64} are useful:
\begin{equation}
\label{eq:translation}
t(\mathbf L)=\exp\left(\frac{i}{\hbar}\mathbf L\cdot\mathbf p-i\frac{\mathbf{\hat z}\cdot(\mathbf L\times\mathbf r)}{\ell^2}
\right),
\end{equation}
where $\mathbf p=\frac{\hbar}{i}\nabla - e\mathbf A$.
In the current gauge, these act as $t(\mathbf L)\psi(\mathbf r) =
\exp(\frac{ix\mathbf{\hat y}\cdot\mathbf L}{\ell^2})\psi(\mathbf r+\mathbf L)$.
We will require each state and the implied density matrix to obey twisted boundary conditions
with twist angles $\phi_{1,2}$,
\begin{equation}
\label{eq:twisted}
t(\mathbf L_{1,2})\psi(\mathbf r)=e^{i\phi_{1,2}}\psi(\mathbf r).
\end{equation}
The two conditions are mutually compatible only if the parallelogram is pierced by an integral number of flux quanta,
\begin{equation}
N_\phi=\frac{\left|\mathbf L_1\times\mathbf L_2\right|}{2\pi\ell^2}
=\frac{L_1 L_2\sin\theta}{2\pi\ell^2}.
\end{equation}
Then the principal domain is also a magnetic unit cell.

If $N_\phi\Re\tau=k$ is an integer, i.e.,
\begin{equation}
\label{eq:restriction}
L_2\cos\theta=\frac{kL_1}{N_\phi},
\end{equation}
straightforward but tedious algebra yields the single-particle density matrix 
\begin{multline}
\label{eq:second}
\rho^\text{PBC}(\mathbf r,\mathbf r';\beta) =
\frac{1}{N_\phi}\rho^\text{open}(\mathbf r,\mathbf r';\beta)\\
\times\sum_{m=0}^{N_\phi-1}\left\{
\vartheta\begin{bmatrix} 0 \\ a_m \end{bmatrix}\left(z_1\Big|\tau_1\right)
\vartheta\begin{bmatrix} 0 \\ 2b_m' \end{bmatrix}(z_2|\tau_2)+\right.\\
\left.+(-1)^k\vartheta\begin{bmatrix} 0 \\ a_m+\frac{1}{2} \end{bmatrix}\left(z_1\Big|\tau_1\right)
\vartheta\begin{bmatrix} \frac{1}{2} \\ 2b_m' \end{bmatrix}(z_2|\tau_2)
\right\},
\end{multline}
where we have factored out $\rho^\text{open}$, the density matrix for open boundary conditions:
\begin{multline}
\label{eq:openbc}
\rho^\text{open}(\mathbf r,\mathbf r';\beta)=\frac{1}{2\pi\ell^2}\frac{\sqrt u}{1-u}\\
\times\exp\left(-\frac{1+u}{1-u}\frac{\left|\mathbf r-\mathbf r'\right|^2}{4\ell^2}+\frac{i(x'-x)(y+y')}{2\ell^2}\right),
\end{multline}
where $\ell=\sqrt{\frac{\hbar}{eB}}$ is the magnetic length, $u=e^{-\beta\hbar\omega_c}$,
and $\omega_c=\frac{eB}{m}$ is the cyclotron frequency \cite{comment3}.
Above, we have used Jacobi elliptic functions with characteristics \cite{Mumford87,comment1}
\begin{equation}
\label{eq:theta}
\vartheta\begin{bmatrix} a \\ b \end{bmatrix}(z|\tau) = \sum_ne^{i\pi\tau(n+a)^2+2i(n+a)(z+b\pi)}.
\end{equation}
The arguments in Eq.~(\ref{eq:second}) are defined as
\begin{equation}
\begin{split}
\tau_1&=\frac{i}{\pi}\left(\frac{L_1}{2\ell N_\phi}\right)^2\frac{1+u}{1-u},\\
z_1&=\frac{L_1}{4\ell^2 N_\phi}\left(y+y' + i(x'-x)\frac{1+u}{1-u}\right),\\
\tau_2&=i\pi\left(\frac{2\ell N_\phi}{L_1}\right)^2\frac{1+u}{1-u},\\
z_2&=\frac{N_\phi\pi}{L_1}\left(x+x' + i(y-y')\frac{1+u}{1-u}\right);
\end{split}
\label{eq:zdef}
\end{equation}
and the constants related to boundary conditions are
\begin{equation}
\label{eq:abdef}
\begin{split}
a_m&=\frac{\phi_1}{2\pi N_\phi} + \frac{m}{N_\phi}, \\
b_m&=-\frac{\phi_2}{2\pi} - \frac{N_\phi\Re\tau}{2}, \\
b_m'&=b_m+N_\phi a_m\Re\tau.
\end{split}
\end{equation}
The derivation of Eq.~(\ref{eq:second}) is delegated to Appendix \ref{app:dm}.

\begin{figure}[htbp]
\begin{center}
\includegraphics[width=0.49\columnwidth]{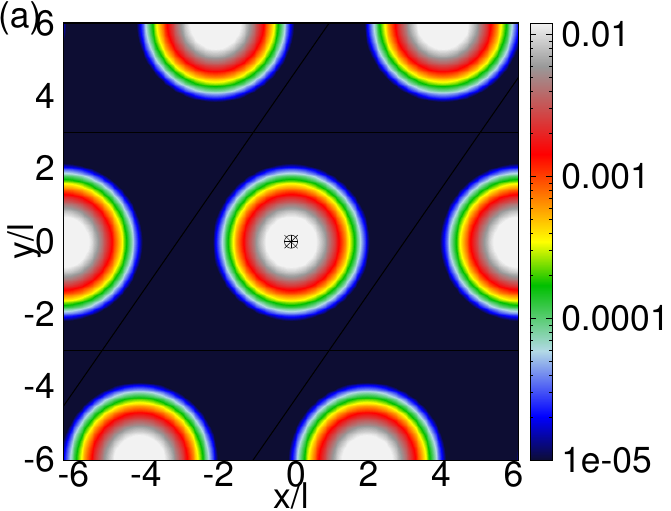}
\includegraphics[width=0.49\columnwidth]{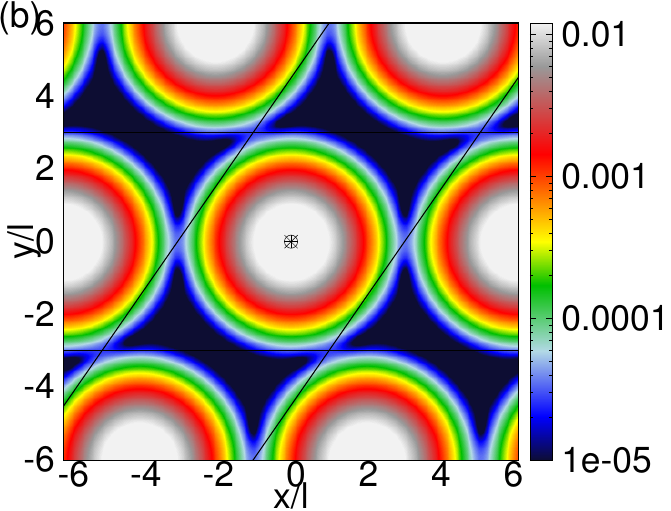}
\includegraphics[width=0.49\columnwidth]{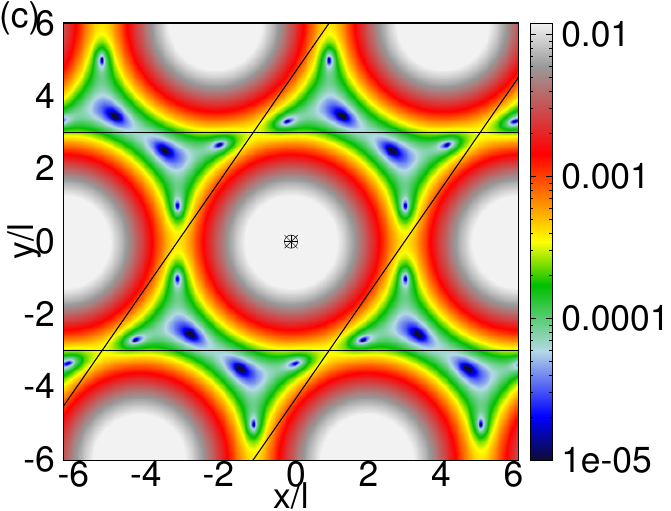}
\includegraphics[width=0.49\columnwidth]{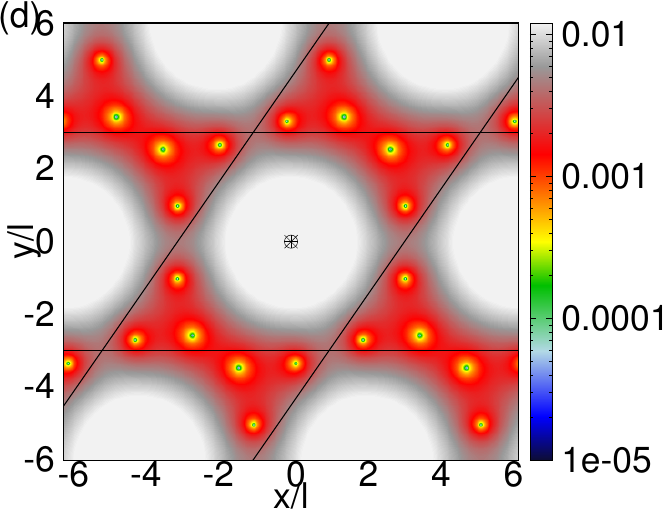}
\end{center}
\caption{\label{fig:tempmove}
The dependence of $|\rho^\text{PBC}(\mathbf r,\mathbf r';\beta)|$ on imaginary time $\beta$.
There are $N_\phi=6$ flux quanta in the principal domain, $L_2/L_1=1.17$, $\theta\approx55^\circ$,
$\phi_1=\phi_2=0$, and we have fixed $\mathbf r'$ at the origin.
The panels correspond to $\beta\hbar\omega_c=0.3$, 0.7, 1.1, and 5, respectively.
}
\end{figure}

The behavior of the density matrix is shown in Fig.~\ref{fig:tempmove} for the
most general case, an oblique unit cell.
For small imaginary time (high temperature) $|\rho^\text{PBC}(\mathbf r,\mathbf r';\beta)|$
has a small Gaussian peak around $\mathbf r'$, which is fixed at the origin in the figure.
This peak spreads out by diffusion as $\beta$ is increased,
and eventually the Gaussians from neighboring unit cells start to overlap appreciably.
However, the density matrix also has a phase due to the external magnetic field,
which gives rise to an interference pattern in this time range.
There is destructive interference at certain points, which effectively arrests the diffusion.
[We will analyze the zeros of $\rho^\text{PBC}(\mathbf r,\mathbf r';\beta)$ below.]
Beyond a certain value of $\beta$, the picture is essentially stationary.

We note that $|\rho^\text{PBC}(\mathbf r,\mathbf r';\beta)|$ is \textit{not} invariant for a simultaneous displacement
of both $\mathbf r$ and $\mathbf r'$ by the same vector $\mathbf d$,
which corresponds to choosing a shifted magnetic unit cell on the plane
for compactification by periodic boundary conditions, except for special choices of $\mathbf d$.
This is understood easily by noting that the second characteristic $b_m$ appears in Eq.~(\ref{eq:theta})
as a simple additive constant to the variable $z$, letting us rewrite Eq.~(\ref{eq:second}) as
\begin{widetext}
\begin{multline}
\rho^\text{PBC}(\mathbf r,\mathbf r';\beta) =
\frac{1}{N_\phi}\rho^\text{open}(\mathbf r,\mathbf r';\beta)\\
\times\sum_{m=0}^{N_\phi-1}
\left\{
\vartheta\begin{bmatrix} 0 \\ a_m + \frac{L_1}{4\pi\ell^2N_\phi}(y+y') \end{bmatrix}
\left(\frac{\pi N_\phi\tau_1'}{L_1}(x'-x)\Big|\tau_1'\right)
\vartheta\begin{bmatrix} 0 \\ 2b_m'+\frac{N_\phi}{L_1}(x+x') \end{bmatrix}
\left(\frac{\pi(y-y')\tau_2}{2L_2\sin\theta}\Big|\tau_2\right)+\right.\\
\left.+(-1)^k\vartheta\begin{bmatrix} 0 \\ a_m+\frac{1}{2} + \frac{L_1}{4\pi\ell^2N_\phi}(y+y') \end{bmatrix}
\left(\frac{\pi N_\phi\tau_1'}{L_1}(x'-x)\Big|\tau_1'\right)
\vartheta\begin{bmatrix} \frac{1}{2} \\ 2b_m'+\frac{N_\phi}{L_1}(x+x') \end{bmatrix}
\left(\frac{\pi(y-y')\tau_2}{2L_2\sin\theta}\Big|\tau_2\right)
\right\}.
\end{multline}
\end{widetext}
Then it is clear that the arguments of the $\vartheta$ functions depend on the coordinate differences only,
and the displacement of the center of mass can be incorporated in the characteristics as
\begin{equation}
b_m\to b_m+\frac{N_\phi}{L_1}d_x,\quad
a_m\to a_m+\frac{L_1}{2\pi\ell^2N_\phi}d_y.
\end{equation}
These in turn correspond to fluxes \cite{Aharonov59,Byers61},
and the shift of the center of mass corresponds to a change in the
twist angles according to Eq.~(\ref{eq:abdef}):
\begin{equation}
\phi_2\to\phi_2-\frac{2\pi N_\phi}{L_1}d_x,\quad
\phi_1\to\phi_1+\frac{L_1}{\ell^2}d_y.
\end{equation}
Thus the twisted boundary conditions in Eq.~(\ref{eq:twisted}), and, consequently, $|\rho^\text{PBC}(\mathbf r,\mathbf r';\beta)|$,
are invariant only if
\begin{equation}
\label{invariant}
\mathbf d = \left(\frac{L_1}{N_\phi}n_1,\frac{2\pi\ell^2}{L_1}n_2\right)
\end{equation}
for integral $n_1$ and $n_2$.

In the $\beta\to0$ limit the density matrix must satisfy
$\rho(\mathbf r,\mathbf r';\beta)\to\delta(\mathbf r-\mathbf r')$, and this holds for
the density matrix appropriate for open boundary conditions in Eq.~(\ref{eq:openbc}).
Using Eq.~(\ref{eq:second}) and the identities of the traditionally defined Jacobi elliptic functions \cite{comment1}
\[
\vartheta_{3,2}(z|\tau)=\sqrt\frac{i}{\tau}\sum_{n=-\infty}^{\infty}(\pm1)^n\exp\left(-\frac{i\pi}{\tau}\left(n+\frac{z}{\pi}\right)^2\right)
\]
one can check that
\begin{multline}
\rho^\text{PBC}(\mathbf r,\mathbf r';\beta\to0) =\sum_{k_1,k_2}
e^{ik_1\phi_1+ik_2\phi_2-\frac{ixk_2L_2\sin\theta}{\ell^2}}\\
\times\delta\left(x-x'-k_1L_1-k_2L_2\cos\theta\right)\\
\times\delta\left(y-y'-k_2L_2\sin\theta\right),
\end{multline}
which complies with the discrete magnetic translation symmetries
\begin{equation}
\begin{split}
t_{\mathbf r}(n\mathbf L_1+m\mathbf L_2)
\rho^\text{PBC}(\mathbf r,\mathbf r';\beta) &= e^{i(n\phi_1+m\phi_2)}\rho^\text{PBC}(\mathbf r,\mathbf r';\beta),\\
t^\ast_{\mathbf r'}(n\mathbf L_1+m\mathbf L_2)
\rho^\text{PBC}(\mathbf r,\mathbf r';\beta)&=e^{-i(n\phi_1+m\phi_2)}\rho^\text{PBC}(\mathbf r,\mathbf r';\beta),
\end{split}
\label{eq:peri}
\end{equation}
which hold for any $\beta$.

\begin{figure}[htbp]
\begin{center}
\includegraphics[width=\columnwidth]{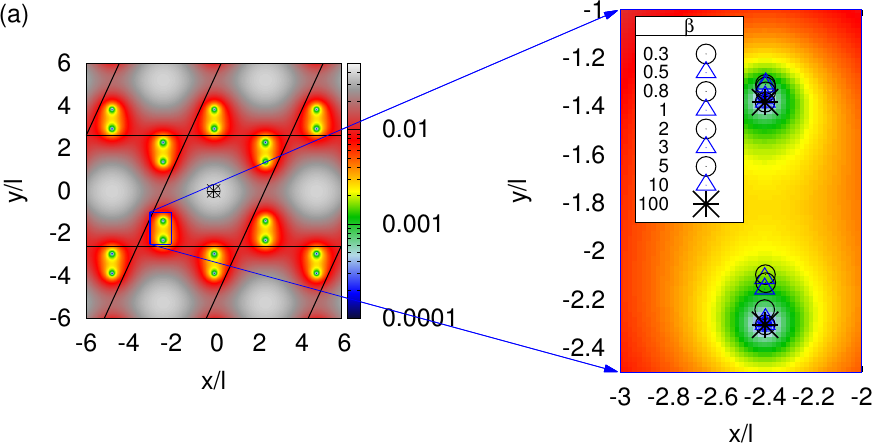}
\includegraphics[width=\columnwidth]{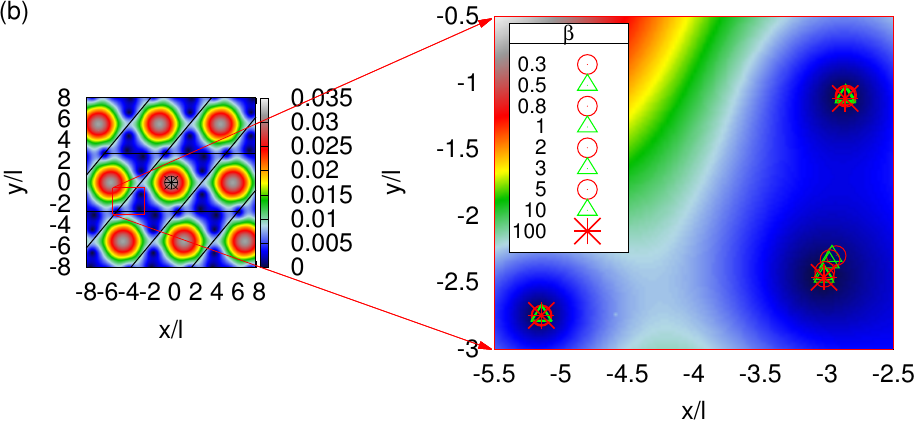}
\end{center}
\caption{\label{zeros}
The low-temperature limit $\beta\to\infty$ of the thermal density matrix.
As no change is discernible beyond $\beta=100$, the density plot has been generated using this value. 
(a) $N_\phi=4$ particles, $\theta\approx65^\circ$, $|\mathbf L_2|/|\mathbf L_1|=1.2$, $\phi_1=\phi_2=0$.
In the zoomed area we show how the zeros move to their asymptotic position as a function of
inverse temperature $\beta$.
(b) The same for $N_\phi=5$, $\theta\approx50^\circ$, $|\mathbf L_2|/|\mathbf L_1|=1.25$, $\phi_1=\phi_2=0$.
Note that one of the zeros is fixed at the corner of the principal region,
which is the generic behavior when $N_\phi$ is odd.
}
\end{figure}

\begin{figure}[htbp]
\begin{center}
\includegraphics[width=0.49\columnwidth]{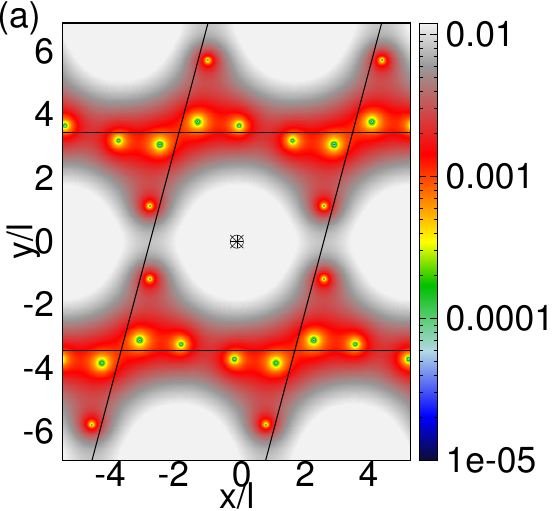}
\includegraphics[width=0.49\columnwidth]{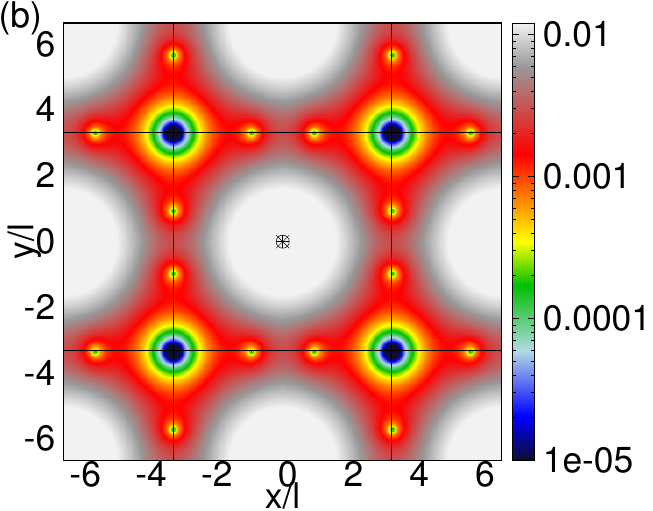}
\includegraphics[width=0.49\columnwidth]{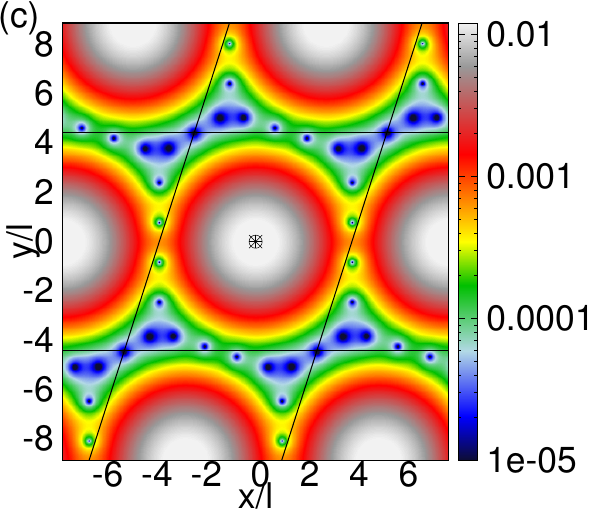}
\includegraphics[width=0.49\columnwidth]{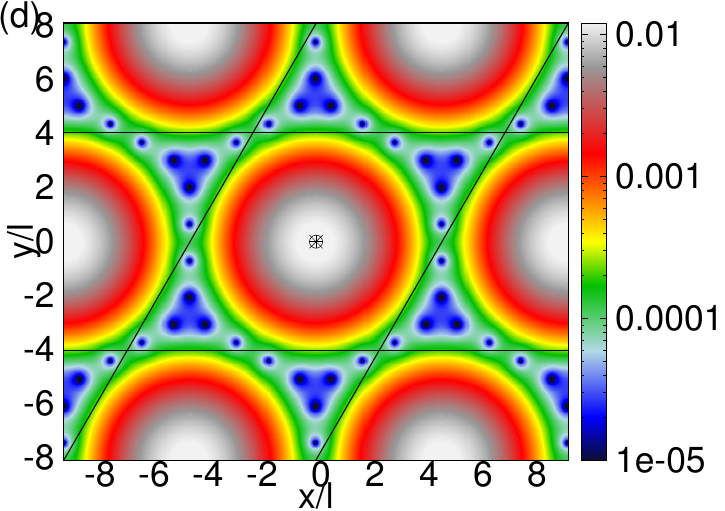}
\end{center}
\caption{\label{zerostruct}
The structure of zeros of the thermal density matrix for $\beta\hbar\omega_c=200$,
where the picture is stationary for different geometries and flux quanta.
We show $|\rho^\text{PBC}(\mathbf r,\mathbf r';\beta)|$, the zeros are the darkest spots.
We set $\phi_1=\phi_2=0$ and fix $\mathbf r'$ at the origin.
(a) Generic torus with $N_\phi=6$, $L_2/L_1=1.13$ and $\theta\approx75^\circ$;
(b) Square principal domain  ($\theta=90^\circ$, $L_2/L_1=1$) with $N_\phi=7$;
(c) Generic torus with $N_\phi=11$, $L_2/L_1=1.19$ and $\theta\approx72^\circ$;
(d) Hexagonal principal domain ($\theta=60^\circ$, $L_2/L_1=1$) with $N_\phi=12$.
}
\end{figure}

In the low-temperature limit, $\beta\to\infty$ ($u\to0$),
the analytic structure of $\rho^\text{PBC}(\mathbf r,\mathbf r';\beta)$ simplifies significantly.
Notice that both for open and periodic boundary conditions, the value of the density matrix goes to
zero at any fixed coordinates $\mathbf r$ and  $\mathbf r'$.
This is an artifact of the zero-point energy $\frac{\hbar\omega_c}{2}$, and it does not appear in averages
as they involve normalization by the partition function $Z(\beta)=\sum_{n=0}^\infty u^{n+1/2}=\frac{\sqrt u}{1-u}$.
We study the analytic structure in the low-temperature limit by factoring out the nonzero factor
$\rho^\text{open}(\mathbf r,\mathbf r';\beta)$ for convenience:
\begin{equation}
\lim_{\beta\to\infty}
\frac{\rho^\text{PBC}(\mathbf r,\mathbf r';\beta)}{\rho^\text{open}(\mathbf r,\mathbf r';\beta)} = f_\infty(z,z'),
\end{equation}
where
\begin{multline}
f_\infty(z,z')=\frac{1}{N_\phi}
\sum_{m=0}^{N_\phi-1}\left\{
\vartheta\begin{bmatrix} 0 \\ a_m \end{bmatrix}\left(\frac{iL_1}{4\ell^2 N_\phi}\left({z'}^\ast-z\right)
\Big|\tilde\tau_1\right)\right.\\
\left.\times
\vartheta\begin{bmatrix} 0 \\ 2b_m' \end{bmatrix}\left(
\frac{N_\phi\pi}{L_1}\left(z+{z'}^\ast\right)
|\tilde\tau_2\right)+\right.\\
\left.+(-1)^k\vartheta\begin{bmatrix} 0 \\ a_m+\frac{1}{2} \end{bmatrix}
\left(\frac{iL_1}{4\ell^2 N_\phi}\left({z'}^\ast-z\right)
\Big|\tilde\tau_1\right)\right.\\
\left.\times\vartheta\begin{bmatrix} \frac{1}{2} \\ 2b_m' \end{bmatrix}
\left(\frac{N_\phi\pi}{L_1}\left(z+{z'}^\ast\right)
|\tilde\tau_2\right)
\right\},
\end{multline}
where
$\tilde\tau_1=\frac{i}{\pi}\left(\frac{L_1}{2\ell N_\phi}\right)^2$
and
$\tilde\tau_2=i\pi\left(\frac{2\ell N_\phi}{L_1}\right)^2$.
$f_\infty(z,z')$ is holomorphic in $z$, and antiholomorphic in $z'$, on the entire complex plane.
Fixing $z'$, the zeros of $f_\infty(z,z')$ can be counted by the argument principle of complex calculus.
Consider the quadruple domain $Q$ with corners $z'+L_1(\pm1\pm\tau)$; cf.\ Fig.~\ref{fig:domain}(b).
We have
\begin{equation} 
\oint_{\partial Q} \frac{d}{dz}\ln\left(f_\infty(z,z')\right) dz=-8\pi iN_\phi,
\end{equation}
which, exploiting the periodicities in Eq.~(\ref{eq:peri}) and the fact that
$\rho^\text{open}(\mathbf r,\mathbf r';\beta)$ is nonzero,
implies that the thermal propagator $\rho^\text{PBC}(\mathbf r,\mathbf r';\beta\to\infty)$
has $N_\phi$ zeros in the principal domain in Fig.~\ref{fig:domain}(a).
At nonzero temperature, the analytic structure of $\rho^\text{PBC}(\mathbf r,\mathbf r';\beta)$ is not simple.
Nevertheless, we have found numerically that the number of zeros in the principal domain is the same
at any finite $\beta$, and the zeros very quickly reach their final location.
See Fig.~\ref{zeros} for illustration.
If $N_\phi$ is odd, there are zeros that do not move at all.
For $\phi_1=\phi_2=0$, in particular, one of them is located in the corners of the principal domain
(which are identical by periodicity).
Fig.~\ref{zerostruct} shows the structure of zeros for different geometries.
Multiple zeros occur in regular cases, as for the square unit cell in panel (b).

\begin{figure*}[htbp]
\begin{center}
\includegraphics[width=0.48\textwidth]{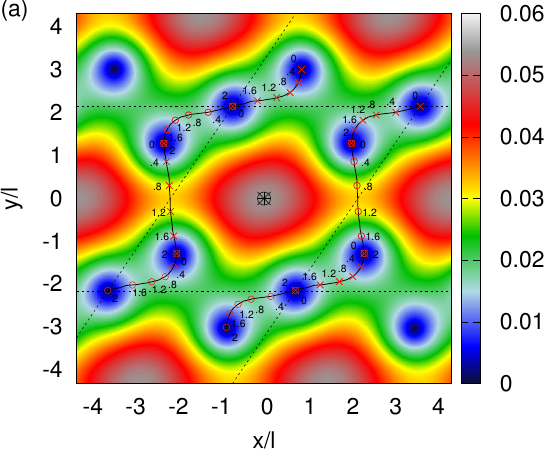}
\includegraphics[width=0.48\textwidth]{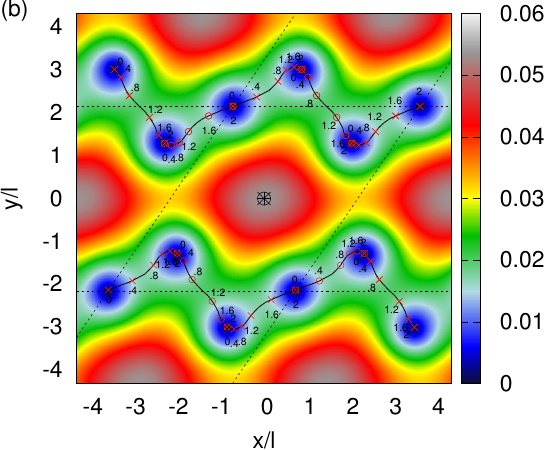}
\end{center}
\caption{\label{zeromove}
The trajectories of the zeros of the thermal density matrix as we tune
(a) the twist angle $\phi_1$  and (b) the twist angle $\phi_2$ between 0 and $2\pi$.
We set $\beta\hbar\omega_c=200$, $\mathbf r'=0$, $N_\phi=2$, $L_2/L_1=1.19$ and $\theta\approx56^\circ$.
The marks at specific points on the trajectories correspond to multiples of $\pi/5$.
The speed of the zeros is not uniform, as visible from the distance between adjacent labeled points.
}
\end{figure*}

In Fig.~\ref{zeromove} we show the motion of the zeros of the thermal density matrix as we tune the twist angles.
Qualitatively, the motion of the zeros shows an interesting analogy with the Hall current:
tuning $\phi_1$ moves them in the $\mathbf L_2$ direction--the direction of the electromotive force
on a charged particle induced by the change of flux--, and conversely.
As a deeper explanation of the motion of the zeros is not crucial to the present work, we leave
the analysis of this issue as an open problem.

\section{Multi-slice sampling}
\label{multislice}

For noninteracting particles and open boundary conditions, the familiar construction of multi-slice
moves \cite{Pollock84} by the bisection method \cite{Ceperley95}
builds a Brownian bridge $R_{L+1},R_{L+2},\dots,R_{R-1}$ between the fixed configurations $R_L$ and $R_R$
at possibly distant slices $L$ and $R=L+2^l$ (mod $M$) on the path.
The deviation of the high-temperature density matrix used in the simulation from the ideal gas case
can be taken into account either at each or just the last level of this recursive procedure.
At each level of this recursive construction we need to know the PDF
of configuration $R_i$, which is to be inserted between $R_{i-s}$ and $R_{i+s}$
at time distances $\pm s\tau$ on the path.
If the ideal gas density matrix $\rho_0(R,R';\beta)$ is \emph{real}, this is simply
\begin{equation}
p(R_i)=\frac{\rho_0(R_{i-s},R_i;s\tau)\rho_0(R_i,R_{i+s};s\tau)}{\rho_0(R_{i-s},R_{i+s};2s\tau)};
\end{equation}
the convolution property in Eq.~(\ref{eq:conv}) ensures that this is a normalized PDF.
If we can sample $p(R_i)$ directly, we implement the \emph{heat-bath rule} for noninteracting particles.
(In fact, with open boundary conditions and zero external magnetic field, $p(R_i)$ is a Gaussian.)
On the other hand, if the free density matrix $\rho_0(R,R',\tau)$ is \emph{complex},
paths must be sampled from the PDF $\prod_{m=1}^{M}|\rho(R_{m-1},R_m;\tau)|$.
As $|\rho_0(R,R',\tau)|$ does not satisfy a convolution property analogous to Eq.~(\ref{eq:conv}),
\begin{equation}
\widetilde p(R_i)=
\frac{|\rho_0(R_{i-s},R_i;s\tau)||\rho_0(R_i,R_{i+s};s\tau)|}{|\rho_0(R_{i-s},R_{i+s};2s\tau)|}
\end{equation}
is not a normalized PDF.
This is not a problem for single-slice moves, but it plagues the bisection method.

First consider how one could adapt multi-slice moves to periodic boundary conditions in the \textit{absence} of a
magnetic field in one dimension.
The single-particle density matrix is \cite{Ceperley95}
\begin{multline}
\rho^\text{PBC}_0(x,x';\beta)=\frac{1}{L}\vartheta_3\left(
\frac{\pi}{L}(x-x') \Big| \frac{4\pi i\lambda\beta}{L^2}\right)=\\
=\frac{1}{\sqrt{4\pi\lambda\beta}}\sum_{n=-\infty}^\infty
\exp\left(-\frac{(x-x'+nL)^2}{4\lambda\beta}\right),
\end{multline}
where $L$ is the period.
(The second equality involves a modular transformation of the function $\vartheta_3(z|\tau)$).
Optimal sampling could be achieved by the heat-bath rule on slice $m$
\begin{equation*}
T^\ast(x'_m|x_{m-1},x_{m+1})=\frac{\rho^\text{PBC}_0(x_{m-1},x_m;\tau)\rho^\text{PBC}_0(x_m,x_{m+1};\tau)}{\rho^\text{PBC}_0(x_{m-1},x_{m+1};2\tau)}.
\end{equation*}
Sampling $x'_m$ from this PDF results in moves that are always accepted for noninteracting particles.
With straightforward algebra,
\begin{multline}
\label{eq:hbath}
T^\ast(x'_m|x_{m-1},x_{m+1})=\\
=\alpha_0\sum_{k=-\infty}^\infty\exp\left(-
\frac{((x_{m+1}+x_{m-1})/2-x'_m+kL)^2}{2\lambda\tau}\right)+\\
+\alpha_1\sum_{k=-\infty}^\infty\exp\left(-
\frac{((x_{m+1}+x_{m-1}+L)/2-x'_m+kL)^2}{2\lambda\tau}\right),
\end{multline}
where
\begin{multline*}
\alpha_i = \frac{1}{\sqrt{2\pi\lambda\tau}}\frac{
\sum_{k'}\exp\left(-\frac{((x_{m+1}-x_{m-1}+iL)/2+k'L)^2}{2\lambda\tau}\right)}
{\sum_{k'}\exp\left(-\frac{(x_{m+1}-x_{m-1}+k'L)^2}{8\lambda\tau}\right)}.
\end{multline*}
$T^\ast(x'_m|x_{m-1},x_{m+1})$ has a very simple structure: the first term
is a collection of the periodic copies of the Gaussian peak centered at $(x_{m+1}+x_{m-1})/2$,
the second term collects peaks at periodic copies of $(x_{m+1}+x_{m-1}+L)/2$.
This suggests a very simple algorithm: with probability
$p=\alpha_0/(\alpha_0+\alpha_1)$ we sample a Gaussian of variance $\lambda\tau$ at
$(x_{m+1}+x_{m-1})/2$, with probability $1-p$ we sample a similar Gaussian at $(x_{m+1}+x_{m-1}+L)/2$.
(With no loss of generality we can choose any of the equivalent peaks,
and map $x'_m$ back to the interval $(-L/2,L/2)$.)
Further, $T^\ast$ in Eq.~(\ref{eq:hbath}) can be applied on any level of the bisection
method to construct a free-particle trajectory between two slices separated by imaginary time $2^l\tau$.
With interactions present, the deviation of the high-temperature density matrix that defines the PDF of paths
from $\rho^\text{PBC}_0$ could be taken into account by a rejection step on the last level of recursion.
(For alternative approaches to periodicity in zero magnetic field, see Ref.~\onlinecite{Cao94}.)

In the presence of an external magnetic field, the density matrix in Eq.~(\ref{eq:second}) is complex-valued. 
We sample paths by the product of the magnitudes of the density matrices that connect
subsequent slices.
If we consider moving a bead $z_m$ on slice $m$ with all other beads fixed.
\begin{equation*}
T^\ast(z'_m|z_{m-1},z_{m+1})=\frac{|\rho^\text{PBC}(z_{m-1},z_m;\tau)||\rho^\text{PBC}(z_m,z_{m+1};\tau)|}{|\rho^\text{PBC}(z_{m-1},z_{m+1};2\tau)|}
\end{equation*}
is not a normalized PDF, but this would not impair the Metropolis algorithm.
As in the $\beta\to0$ limit $|\rho^\text{PBC}(z,z';\tau)|$ with fixed $z'$ tends to a system of Gaussian peaks
centered at $z'+nL_1+mL_1\tau$, just like in the nonmagnetic case, we try the following.
We choose the \textit{a priori} sampling PDF $T(z'_m|z_{m-1},z_{m+1})$ as a collection of four Gaussian
peaks centered at
\begin{equation}
\begin{split}
Z^{z_{m-1},z_{m+1}}_0&=(z_{m-1}+z_{m+1})/2,\\
Z^{z_{m-1},z_{m+1}}_1&=(z_{m-1}+z_{m+1}+L_1)/2,\\
Z^{z_{m-1},z_{m+1}}_2&=(z_{m-1}+z_{m+1}+L_1\tau)/2,\\
Z^{z_{m-1},z_{m+1}}_3&=(z_{m-1}+z_{m+1}+L_1(1+\tau))/2.
\end{split}
\end{equation}
The height of these peaks is proportional to
\begin{equation}
\alpha_i=\frac{|\rho^\text{PBC}(z_{m-1},Z_i;\tau)||\rho^\text{PBC}(Z_i,z_{m+1};\tau)|}
{|\rho^\text{PBC}(z_{m-1},z_{m+1};2\tau)|},
\end{equation}
for $0\le i\le 3$.
We choose peak $i$ with probability $p_i=\alpha_i/(\sum_{j=0}^3\alpha_j)$.
We take into account the fact that the diffusive motion described by both
$|\rho^\text{open}(R,R',\tau)|$ and $|\rho^\text{PBC}(R,R',\tau)|$ is different
from the diffusion in the absence of magnetic field.
Thus the sampled Gaussian has variance $\frac{1-u}{1+u}\ell^2$ with $u=e^{-\hbar\omega_c\tau}$.
Notice that $\frac{1-u}{1+u}\ell^2<\lambda\tau$.

As the heat-bath rule is not obeyed, the acceptance probability is less than unity even
for noninteracting particles in single-slice moves:
\begin{multline}
A(z_m\to z'_m) =
\frac{|\rho^\text{PBC}(z_{m-1},z'_m;\tau)||\rho^\text{PBC}(z'_m,z_{m+1};\tau)}{
|\rho^\text{PBC}(z_{m-1},z_m;\tau)||\rho^\text{PBC}(z_m,z_{m+1};\tau)}\\
\times\frac{T(z_m|z_{m-1},z_{m+1})}{T(z'_m|z_{m-1},z_{m+1})}.
\end{multline}

For multi-slice moves, we proceed as follows.

(i) A trial path is constructed recursively between slices $L$ and $R=L+2^l$.
Midway between slices $L$ and $R$, we choose $z'_{(L+R)/2}$
from one of four Gaussian peaks at $Z^{z_{L},z_{R}}_i$ of
variance $\frac{1-u_1}{1+u_1}\ell^2$, where $u_1=e^{-\hbar\omega_c\tau_1}$ and $\tau_1=2^{l-1}\tau$.
Then we sample $z'_{L+2^{l-2}}$ from one of four Gaussian peaks at $Z^{z_{L},z'_{(L+R)/2}}_i$
and $z'_{R-2^{l-2}}$ from one of four Gaussian peaks at $Z^{z'_{(L+R)/2},z_R}_i$,
all having variance $\frac{1-u_2}{1+u_2}\ell^2$, where $u_2=e^{-\hbar\omega_c\tau_2}$ and $\tau_2=2^{l-2}\tau$.
We continue on subsequent levels, until the trial path $z'_{L+1},\dots z'_{R-1}$ is complete.
During this construction, the ratio of the \textit{a priori} sampling PDFs
\begin{equation}
P_1=\frac{T(z_{L+1},\dots z_{R-1}|z_{L},z_{R})}{T(z'_{L+1},\dots z'_{R-1}|z_{L},z_{R})}
\end{equation}
is stored.

(ii) Once the trial path is available, the ratio of the PDF of the new and the old paths is calculated,
\begin{equation}
P_2=\frac{\prod_{m=L+1}^{R}|\rho(z'_{m-1},z'_m;\tau)|}{
\prod_{m=L+1}^{R}|\rho(z_{m-1},z_m;\tau)|}.
\end{equation}
The constructed trial path is then accepted with probability $A(z\to z')=P_1P_2$.

\begin{figure}[htbp]
\begin{center}
\includegraphics[width=\columnwidth]{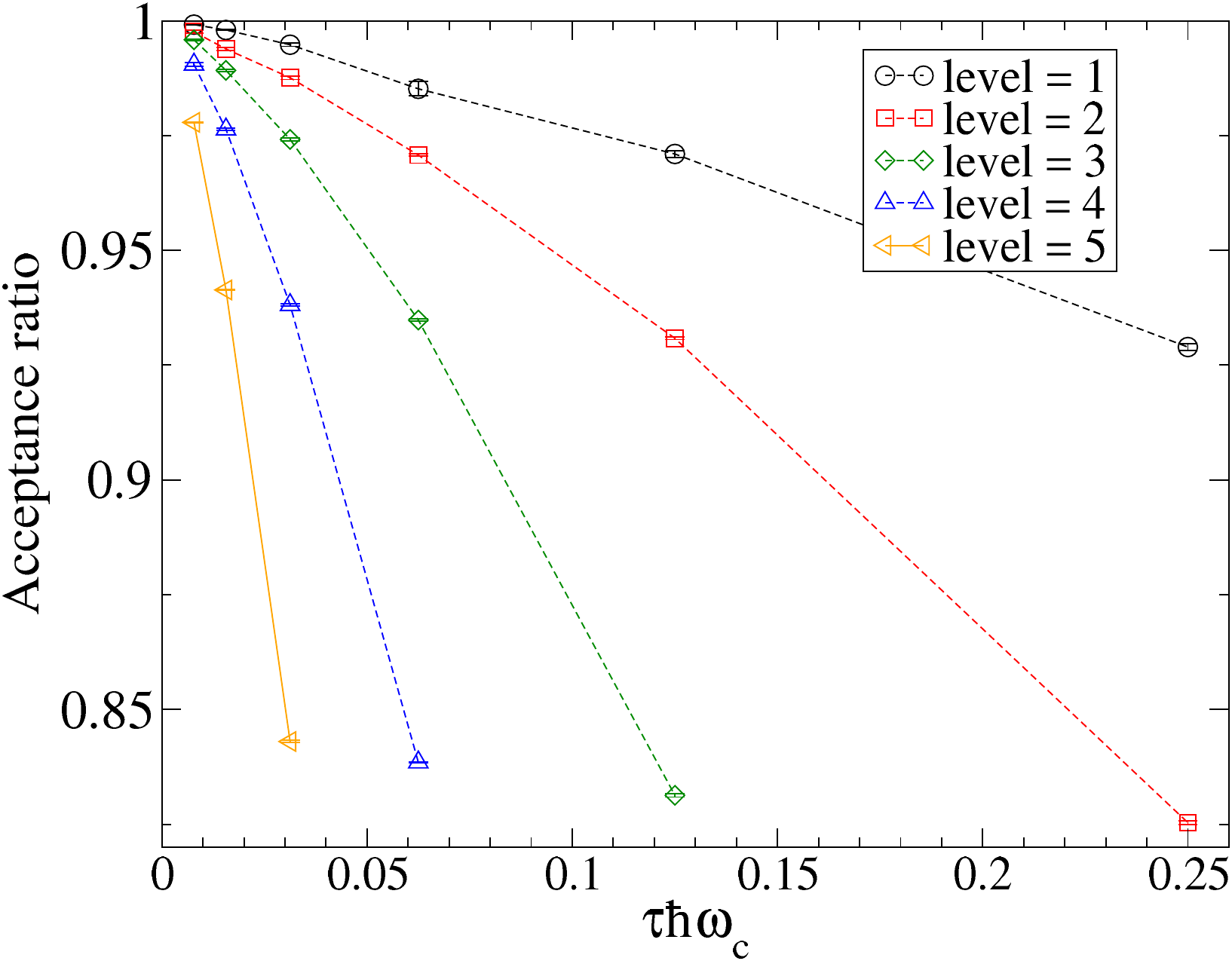}
\end{center}
\caption{\label{fig:stest}
Acceptance ratios for sampling the motion of a single particle on a rectangular
torus pierced by $N_\phi=2$ flux quanta.
The inverse temperature of the system is $\beta\hbar\omega_c=2$,
and the number of slices ranged between $M=8$ and 256.
Level $l$ means that $2^l-1$ slices are updated in each multi-slice move.
}
\end{figure}

For testing the efficiency of the above algorithm, in Fig.~\ref{fig:stest} we show
the acceptance ratio for the simplest possible case, namely the simulation of a single free particle on the torus.
The phase was fixed to the density matrix in Eq.~(\ref{eq:second});
we set $\beta\hbar\omega_c=2$, and there are $N_\phi=2$ flux quanta through a rectangular torus.
(For the computational advantage of choosing $N_\phi$ even, see Appendix \ref{app:comput}.)
For $N$ particles, the acceptance ratio is roughly raised to the $N$-th power; 
this is the baseline that interactions are expected to reduce further.
We have checked systematically that the acceptance ratio depends only weakly on the aspect ratio
or the twist angles.

\section{Application: rotating Yukawa gases}
\label{yukawa}

We consider particles that interact by a repulsive modified-Bessel-function interaction.
The system rotates about the $z$-axis with angular velocity $\Omega$.
In the co-rotating frame it is described by the Hamiltonian
\begin{equation}
\label{eq:corot}
\mathcal H=-\frac{\hbar^2}{2m}\sum_{i=1}^N
\left(\nabla_i - \frac{im}{\hbar}\mathbf\Omega\times\mathbf r\right)^2
+\epsilon\sum_{i<j}K_0\left(\frac{r_{ij}}{a}\right),
\end{equation}
where $\epsilon$ and $a$ characterize the strength and the range of the interaction, respectively.
The correspondence between $\Omega$ and the formerly defined cyclotron frequency and magnetic length scales is
\begin{equation}
\omega_c=2\Omega\quad\text{and}\quad\ell=\sqrt{\frac{\hbar}{2m\Omega}}.
\end{equation}
We consider both Bose and spinless Fermi systems.

In cold atomic experiments a confinement potential is also present, which is weakened by the centrifugal
force in the co-rotating frame.
We do not include these terms; we describe a homogeneous portion of the gas.
As is apparent from Eq.~(\ref{eq:corot}), the Coriolis force couples to momenta just like
a uniform magnetic field does for charged particles \cite{Wilkin98,Cooper08}.

For $\Omega=0$, a mathematically equivalent system arises in type-II superconductors,
where the bosons correspond to Abrikosov vortex lines \cite{Nelson89}.
Both the ground state \cite{Magro93} and the finite-temperature \cite{Nordborg97} phase diagram
of this time-reversal invariant system have been explored by quantum Monte Carlo techniques.

There are four energy scales in the problem: the temperature $k_\text{B}T\equiv\beta^{-1}$,
the cyclotron energy $\hbar\omega_c$, the interaction strength $\epsilon$, and
the energy that corresponds to the interaction length scale, $\hbar^2/(2ma^2)$.
We introduce the dimensionless parameters
\begin{equation}
\begin{split}
\beta^\ast = \beta\hbar\omega_c=2\beta\hbar\Omega,\quad\quad
\rho^\ast=\rho a^2,\\
\Lambda=\sqrt{\frac{\hbar^2}{2ma^2\epsilon}},\quad\quad
\kappa=\frac{a}{\ell}=a\sqrt{\frac{2m\Omega}{\hbar}},
\end{split}
\end{equation}
where $\rho$ is the particle density and $\ell$ is the magnetic length.
$\Lambda$ is the de Boer interaction strength parameter.
We could also have used
\begin{equation}
\tilde\beta=\beta\epsilon
\end{equation}
to turn the inverse temperature dimensionless;
the two dimensionless temperature parameters are related as $\tilde\beta=\beta^\ast/(2\kappa^2\Lambda)$.
The dimensionless density can be related to the filling factor $\nu$ of Landau levels as $\rho^\ast=\kappa^2\nu/2\pi$.

With time-reversal symmetry, the system orders in a triangular lattice for strong interaction (small $\Lambda$)
\cite{Magro93,Nordborg97}.
With this prior knowledge, we choose the aspect ratio of the rectangular simulation cell so that it can
accommodate a finite piece of a triangular lattice with periodic boundary conditions.
This means $\sqrt3/2$ for $N=4$, 12 and 16 particles, and $\sqrt3$ for $N=8$ particles.
We emphasize that this choice is the only \textit{a priori} input to our simulation.
The ideal Bose and Fermi gas, respectively, that we use for phase fixing is not ideal
either for a crystal or a correlated liquid.

\begin{figure*}[htbp]
\begin{center}
\includegraphics[width=0.32\textwidth, keepaspectratio]{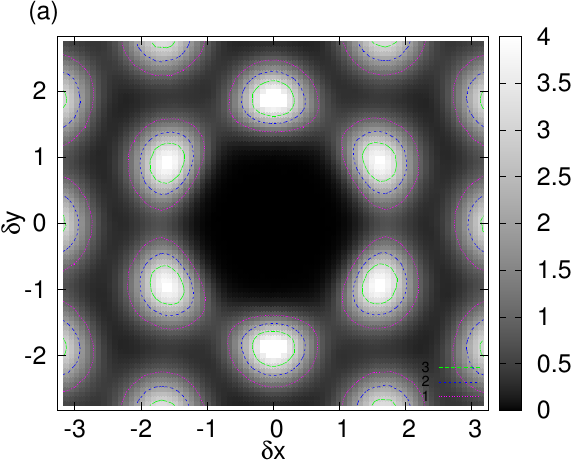}
\includegraphics[width=0.32\textwidth, keepaspectratio]{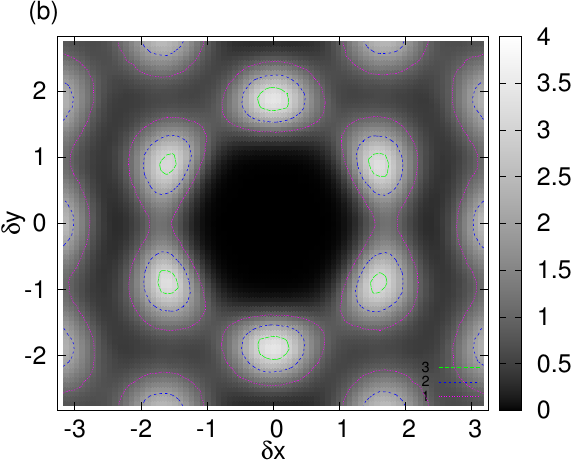}
\includegraphics[width=0.32\textwidth, keepaspectratio]{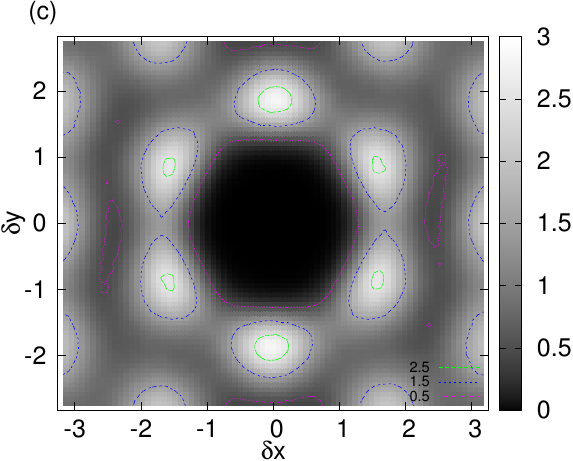}

\vspace*{0.1cm}
\includegraphics[width=0.32\textwidth, keepaspectratio]{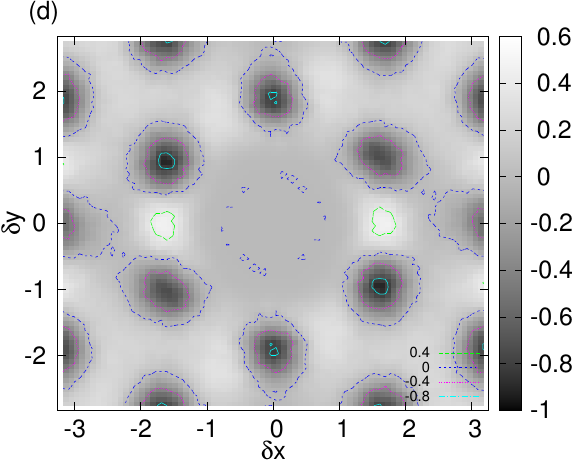}
\includegraphics[width=0.32\textwidth, keepaspectratio]{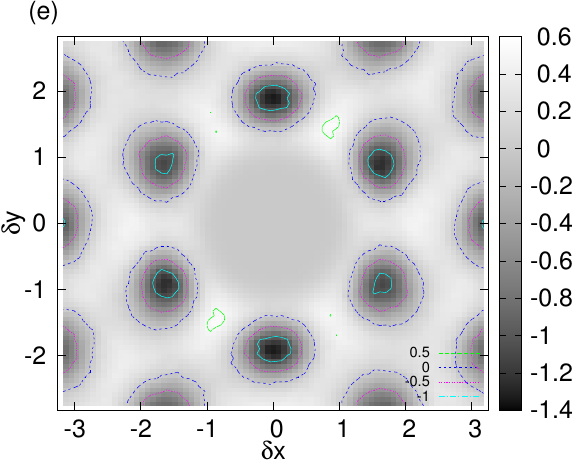}
\end{center}
\caption{\label{fig:bose}
(a-c) The pair-correlation function for $N=12$ bosons at density $\rho a^2=0.02$ ($\kappa=0.25066$),
at filling factor $\nu=2$ (i.e., $N_\phi=6$ flux quanta piercing the torus) and interaction strength $\Lambda=0.035$,
0.04, and 0.045, respectively.
$M=32$ slices were used, the imaginary time-step is $\tau=0.015625$.
The temperature is low on the scale of interactions, as $\tilde\beta=114$, 99, and 88 in panels (a) to (c).
Panels (d) and (e) show the differences of the pair-correlation functions,
$g_{\Lambda=0.04}-g_{\Lambda=0.035}$ and $g_{\Lambda=0.045}-g_{\Lambda=0.04}$, respectively,
as $\Lambda$ is changed for systems shown in the top row.
The triangular lattice of dark spots shows the decreasing crystalline correlation as $\Lambda$ is increased.
The small deviations from perfect $C_6$ symmetry in panels (b) and (c)
can be attributed to imperfect thermalization, and could be reduced by longer Monte Carlo runs.
Taking the differences between pair-correlation functions in panels (d) and (e) amplifies these small errors.
}
\end{figure*}

In analogy to free-particle nodes, we fix the phase to the density matrix of the ideal gas,
\begin{equation}
\label{eq:freefermi}
\rho_F(R,R';\beta) = \text{Det}(\rho^\text{PBC}(\mathbf r_i,\mathbf r'_j;\beta))
\end{equation}
for fermions, and
\begin{equation}
\label{eq:freebose}
\rho_B(R,R';\beta) = \text{Perm}(\rho^\text{PBC}(\mathbf r_i,\mathbf r'_j;\beta))
\end{equation}
for bosons; Perm stands for the permanent.
As we will see, such an ansatz is sufficiently nonrestrictive for reasonable predictions \cite{comment2}.
(Computationally, of course, the Fermi case is easier.)
As phase-fixing for PIMC has already been discussed in the literature \cite{Akkineni08},
we are content with summarizing the technicalities in Appendix \ref{app:pfaction}.

\begin{figure*}[htbp]
\begin{center}
\includegraphics[width=0.32\textwidth, keepaspectratio]{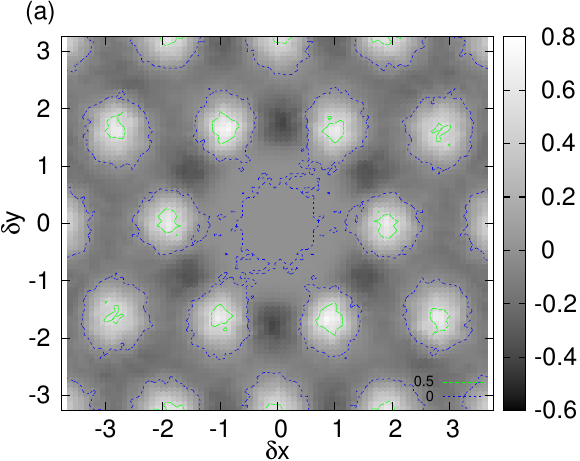}
\includegraphics[width=0.32\textwidth, keepaspectratio]{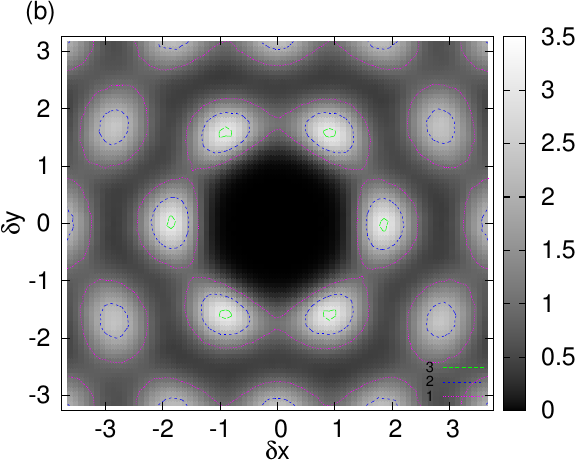}
\includegraphics[width=0.32\textwidth, keepaspectratio]{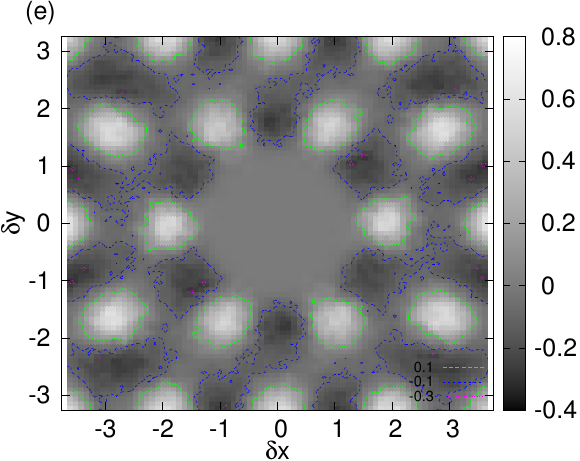}

\vspace*{0.1cm}
\includegraphics[width=0.32\textwidth, keepaspectratio]{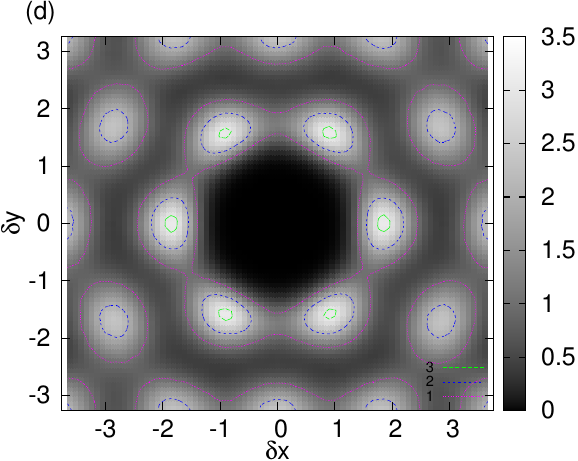}
\includegraphics[width=0.32\textwidth, keepaspectratio]{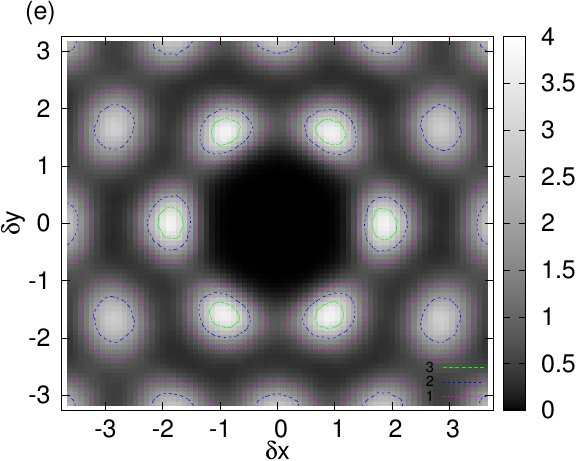}
\includegraphics[width=0.32\textwidth, keepaspectratio]{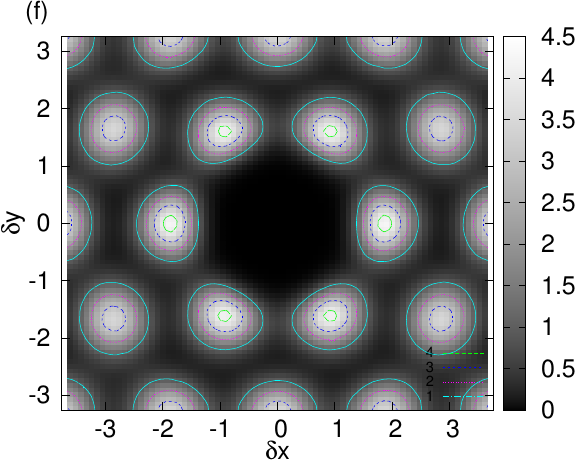}

\vspace*{0.1cm}
\includegraphics[width=0.32\textwidth, keepaspectratio]{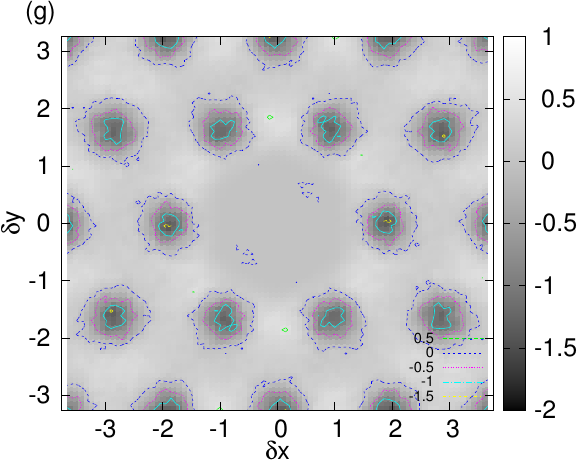}
\includegraphics[width=0.32\textwidth, keepaspectratio]{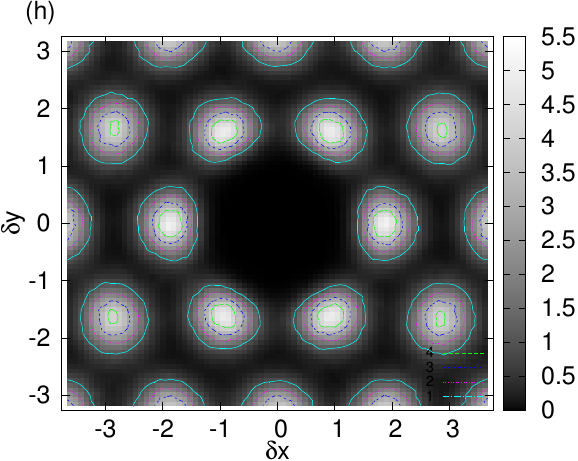}
\includegraphics[width=0.32\textwidth, keepaspectratio]{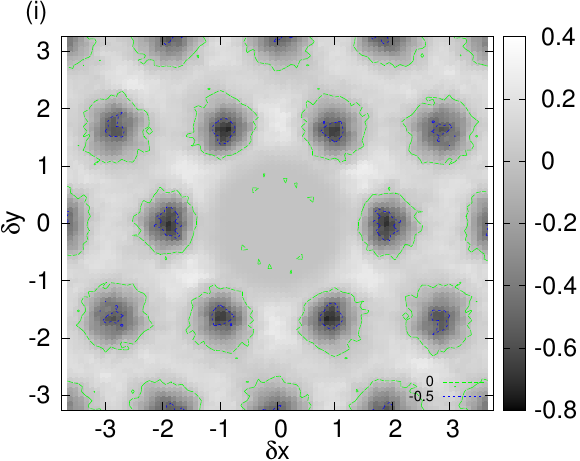}
\end{center}
\caption{\label{fig:fermi}
Second row (d)-(f): the pair-correlation function for $N=16$ fermions at density $\rho a^2=0.02$ ($\kappa=0.25066$),
at filling factor $\nu=2$ (i.e., $N_\phi=8$ flux quanta piercing the torus) and interaction strength $\Lambda=0.035$
at inverse temperature $\beta^\ast=0.4$, 0.5, and 0.6, respectively.
In (d) $M=16$ slices were used, $\tau=0.025$ and $\tilde\beta=132$;
in (e) $M=16$, $\tau=0.03125$ and $\tilde\beta=114$;
in (f) $M=24$, $\tau=0.025$ and $\tilde\beta=99$.
Panels (a) and (c) show the differences of the pair-correlation functions
$g_{\beta=0.4}-g_{\beta=0.5}$ and $g_{\beta=0.6}-g_{\beta=0.5}$, respectively,
between colder and warmer systems shown in consecutive panels in the second row.
The triangular lattice of bright spots shows the increasing crystalline correlation as the temperature is decreased.
Second column (b), (e), and (h): the pair-correlation function as the temperature is held fixed at $\beta^\ast=0.5$, but the
de Boer parameter is tuned from $\Lambda=0.03$ in panel (h) to $\Lambda=0.04$ in panel (b).
Panels (g) and (i) show the differences of the pair-correlation functions
$g_{\Lambda=0.035}-g_{\Lambda=0.03}$ and $g_{\Lambda=0.04}-g_{\Lambda=0.035}$, respectively,
as $\Lambda$ is tuned for systems shown in the second column.
The triangular lattice of dark spots shows the decreasing crystalline correlation as $\Lambda$ is increased.
}
\end{figure*}

The pair-correlation function for $N=12$ bosons at $\beta^\ast=0.5$ is shown in Fig.~\ref{fig:bose}.
Qualitatively, the transition to the crystalline structure is captured.
Due to computational limitations, however, we cannot simulate more than 12 bosons.
The pair-correlation for a larger Fermi system is shown in Fig.~\ref{fig:fermi}.
The qualitative behavior is similar.
Notice that the small $\beta^\ast$ means that while temperature destroys magnetic effects,
it is still small on the interaction energy scale; $\tilde\beta$ is on the scale of $10^2$.
(In the absence of flux, Ref.~\cite{Nordborg97} finds essentially ground-state behavior at $\tilde\beta\approx300$.)

It is customary to characterize the crystalline order by the Lindemann ratio
$\gamma = \sqrt{\frac{1}{N}\sum_{i=1}^N\left\langle (\mathbf r_i - \mathbf R_i)^2\right\rangle}/d$,
where $d$ is the lattice constant and $\mathbf R_i$ is the lattice point nearest to particle $i$.
In our case, however, we cannot hold the center of mass fixed during Monte Carlo, because the
simultaneous shift of all beads by the same vector is not a symmetry,
except for some discrete values, as discussed in Sec.~\ref{sec:freedm}.
One could locate the lattice points with reference to the instantaneous center of mass
assuming the lattice is triangular with the lattice constant implied by the density.
But this procedure underestimates $\gamma$.
Hence, we decided to infer the qualitative behavior from the pair-correlation function instead.

\begin{figure}[htbp]
\begin{center}
\includegraphics[width=\columnwidth, keepaspectratio]{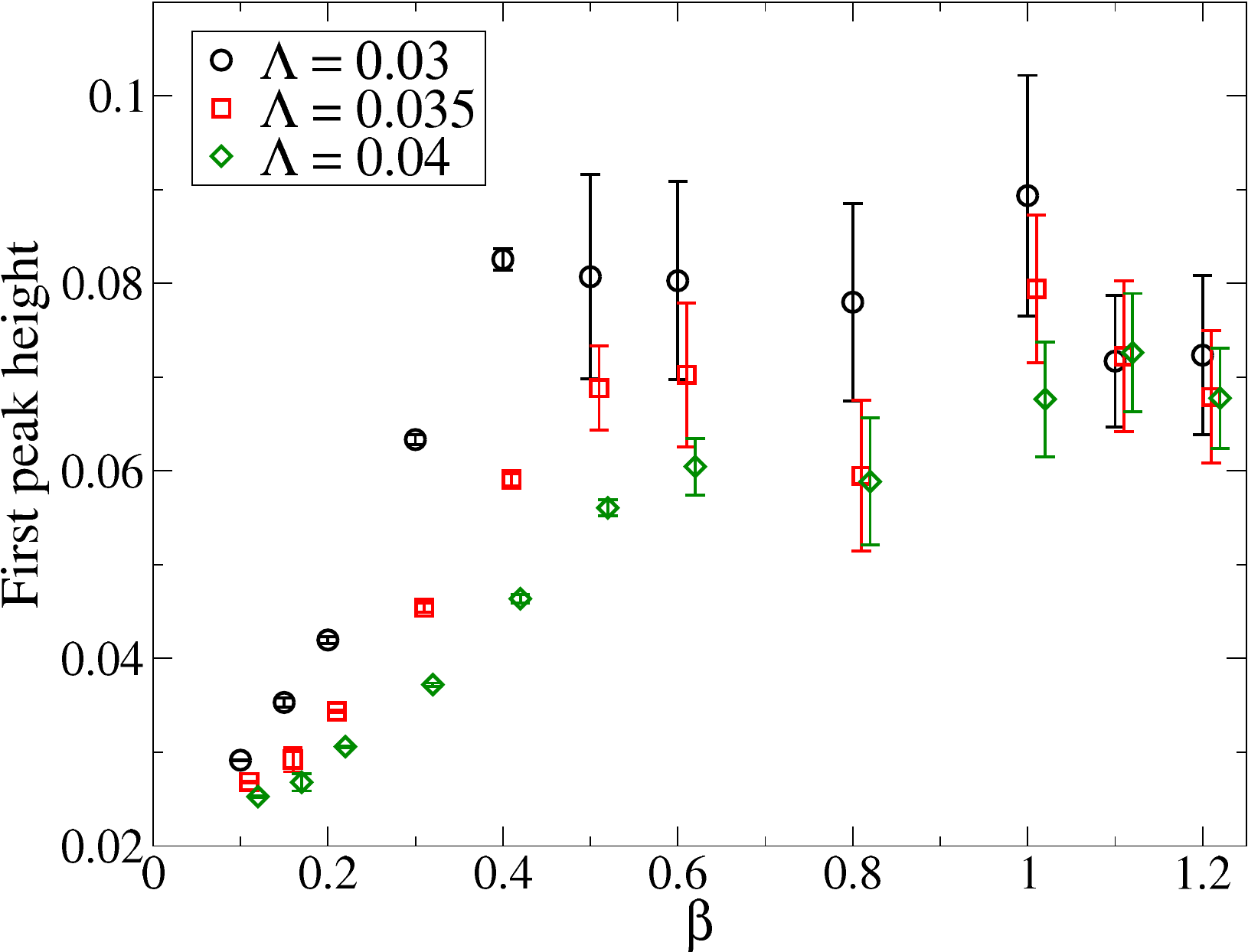}
\end{center}
\caption{\label{fig:nonmon2}
The evolution of the first peak of the pair-correlation function for $N=16$ fermions at flux $N_\phi=8$
at density $\rho a^2=0.02$ ($\kappa=0.25066)$, as a function of the inverse temperature
for some values of the de Boer parameter $\Lambda$ for which crystalline structure is manifest
at intermediate temperatures.
A small horizontal shift has been applied to the last two curves to make the overlapping error bars visible.
}
\end{figure}

By inspecting the difference of the pair-correlation functions of systems that differ only by one
parameter, we have checked that in the $\beta^\ast<1$ range our method reproduces the
tendencies known for the nonrotating system:
the crystalline tendency becomes stronger with increasing $\beta^\ast$ at fixed $\Lambda$ and $\rho^\ast$,
as seen in the related panels of Figs.~\ref{fig:bose}, \ref{fig:fermi} and \ref{fig:nonmon2}, and
it becomes stronger when decreasing $\Lambda$ at fixed $\beta^\ast$ and $\rho^\ast$.
Also, Fermi systems show stronger peaks in the pair-correlation than Bose systems at identical temperature,
density, and de Boer parameter $\Lambda$.
It is not possible to go beyond qualitative statements now, as neither finite-size scaling nor a
$\tau\to0$ extrapolation has been performed.
With the prior knowledge that the melting transition is first-order, it will be necessary to
perform simulations with the particle density as a dynamical variable \cite{Nordborg97}. 
As our goal is to demonstrate the applicability of PIMC to bulk systems in the absence of
time-reversal symmetry, and not an in-depth analysis of the Yukawa system,
we delegate such a quantitative analysis to future work.

\begin{figure}[htbp]
\begin{center}
\includegraphics[width=\columnwidth, keepaspectratio]{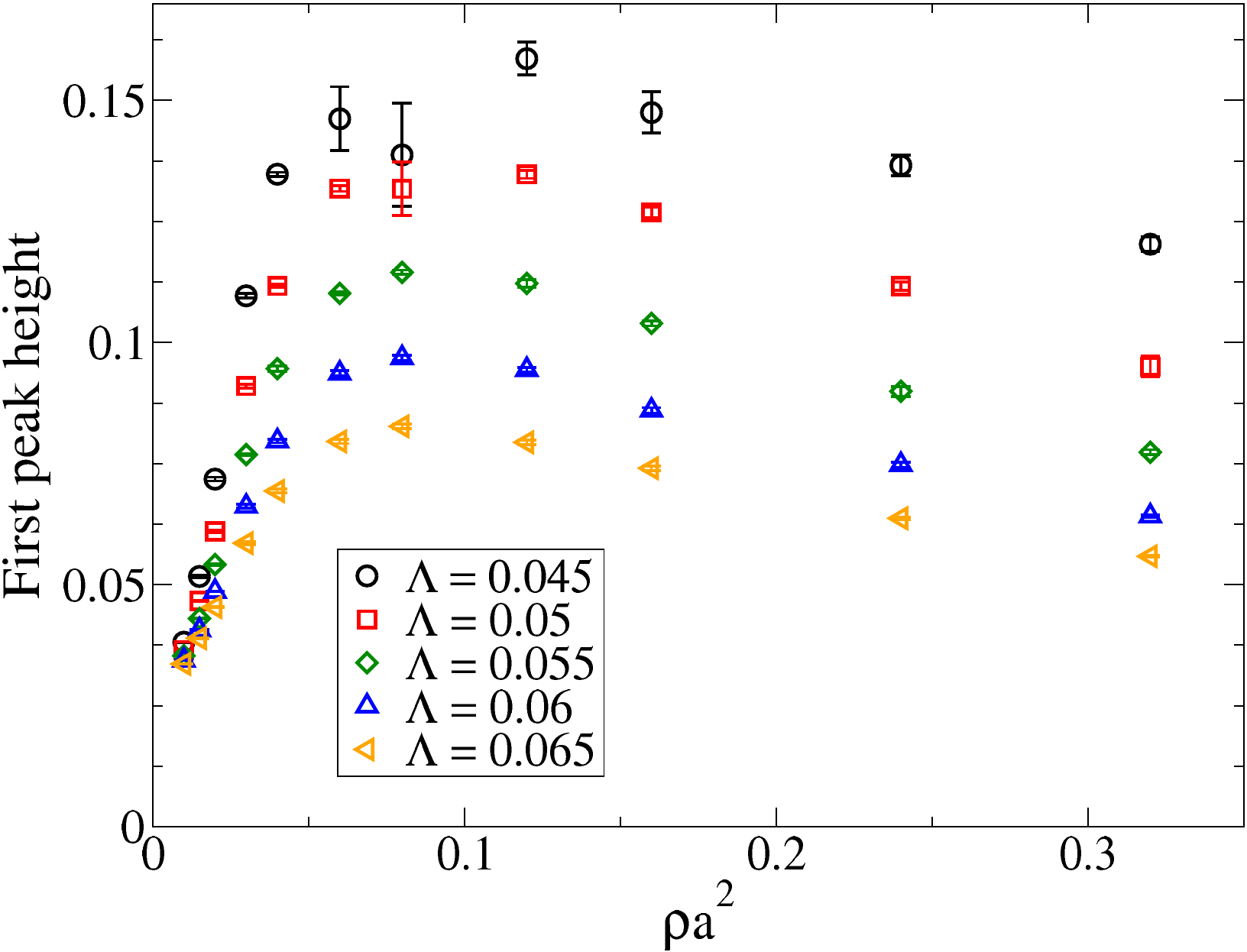}
\end{center}
\caption{\label{fig:density}
The height of the first peak of the pair-correlation function for $N=12$ fermions at $\beta^\ast=0.5$ and
$N_\phi=6$, for various $\Lambda$ de Boer parameter values as the function of density $\rho a^2$.
The nonmonotonic evolution indicates that crystalline order exists only for a
limited range of densities.}
\end{figure}

For $\Omega=0$, Yukawa bosons are known to exhibit nonmonotonic behavior as a function of density:
at fixed interaction strength $\Lambda$ the system first crystallizes with increasing density,
then at sufficiently high density it melts again.
Due to computational limitations, we have only been able to verify this for the Fermi system.
Fig.~\ref{fig:density} shows the evolution of the first peak of the pair-correlation function as the density changes
at fixed $\beta^\ast$ and $\Lambda$ values for fermions.
Apparently, crystalline order prevails only for intermediate densities, just like for bosons
at zero temperature in the absence of rotation \cite{Magro93}.
Determining the phase boundary will require more extensive simulations.

In the $\beta^\ast>1$ range the strength of the crystalline correlations apparently
starts to weaken as a function of the inverse temperature for fermions.
Such an evolution is shown in Fig.~\ref{fig:nonmon2} for various de Boer interaction
parameters $\Lambda$ as the temperature is tuned from $\beta^\ast=0.1$ to 1.2.
The pair correlation becomes more crystalline in the $\beta^\ast\lesssim0.6$ range,
then stagnates, and seems to weaken again above $\beta^\ast\approx1$.
Clearly, more comprehensive calculations in the large-$\beta$ region are necessary to ascertain that
this tendency is robust.
If so, it indicates the competition of the homogeneous integer quantum Hall liquid state (the ground state
candidate for this particular density) and the density-wave ordering,
which requires thermal excitations above the cyclotron gap that the interaction can organize in a crystalline order.
This competition is, of course, not expected for bosons or bolzmannons;
for the latter we have checked the monotonic evolution up to $\beta^\ast=1.8$.

\begin{figure}[htbp]
\begin{center}
\includegraphics[width=\columnwidth, keepaspectratio]{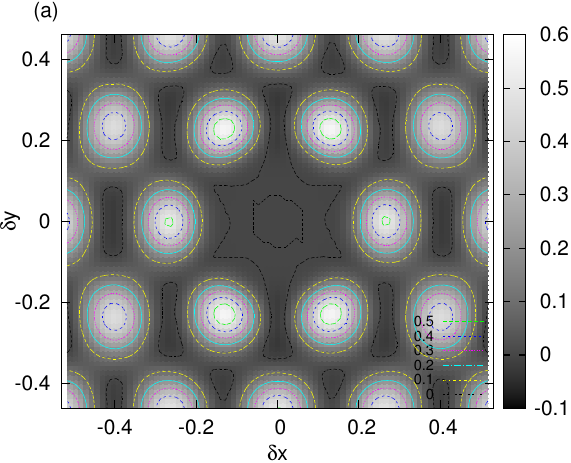}

\includegraphics[width=\columnwidth, keepaspectratio]{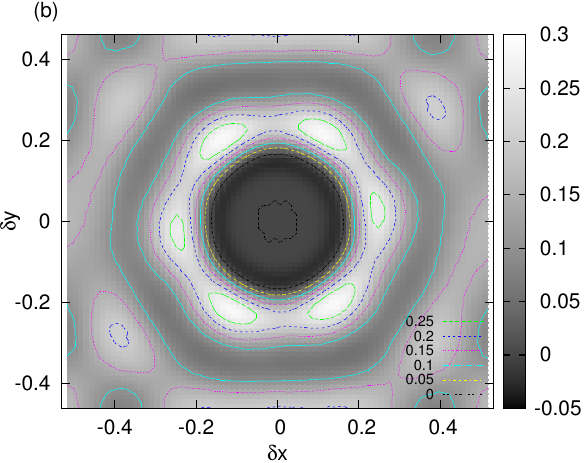}
\end{center}
\caption{\label{fig:flux}
The difference $g_{N_\phi=2}(\mathbf r)-g_{N_\phi=6}(\mathbf r)$ between pair-correlation functions at
different flux densities (6 and 2 flux quanta through the torus) at $\beta^\ast=0.5$ for $N=16$ and
$\rho a^2=0.02$ for fermions (a) and bolzmannons (b).
The total area of the simulation call is scaled to unity, thus the peak locations may coincide.
(The filling factor corresponding to $N_\phi=6,2$ is $\nu=\frac{8}{3},8$, respectively.)
The triangular lattice of bright peaks correspond to stronger crystalline correlations at smaller flux density.
Bolzmannons in panel (b) are still liquid-like; a small rotation of the hexagonally distorted
rings from directions where crystalline structure will emerge can be attributed to imperfect thermalization.
}
\end{figure}

It is also interesting to review the evolution of the pair-correlation as a function
of flux density (magnetic field or Coriolis-force) when the particle density $\rho^\ast$ is held fixed.
Again, we could study this only for fermions and bolzmannons; some of the results are shown in Fig.~\ref{fig:flux}.
(Notice that while $\beta^\ast$ is kept constant, the system becomes colder on the interaction
energy scale as $\tilde\beta=\beta^\ast\nu/(4\pi\Lambda\rho^\ast)$ with $\nu=N/N_\phi$;
the ratio of the interaction and the magnetic length scale also changes as $\kappa=\sqrt{2\pi\rho^\ast/\nu}$.)
We see that the system becomes more crystalline as the number of flux quanta is decreased,
which is only possible in very crude steps with $N=16$, the largest system we simulated routinely.
The tendency is qualitatively the same for fermions and bolzmannons, but it is stronger for fermions.
Note that the flux density would localize particles on the scale of the magnetic length,
which is greater than the lattice constant for $\kappa<1$.
On the other hand, it is more difficult to obtain converged results for smaller flux densities,
which is no doubt related to the shortening of the length scale on which the change of the
phase of the many-body wave function can be considered smooth for the phase-fixing procedure;
in the limit of vanishing magnetic field, we approach the sudden sign changes that
are treated by node fixing in time-reversal-symmetric simulations.

We note that the PIMC calculations for $N=12$ bosons in Fig.~\ref{fig:bose} required about one day of thermalization
and two days of data collection on a single Intel Xeon X5660 CPU core at 2.8 GHz,
while the calculations for $N=16$ fermions in Fig.~\ref{fig:fermi} were about half that long.
With increasing inverse temperature the number of slices also has to be increased; the most expensive
calculation we performed was for $\beta=1.1$ in Fig.~\ref{fig:nonmon2}, with three days of thermalization and
eleven days of data collection.
The number of flux quanta hardly affects the resources needed: each of the calculations compared in Fig.~\ref{fig:flux}(a)
required about three plus six days; the calculations for distinguishable particles in Fig.~\ref{fig:flux}(b) were about
a factor of 3 cheaper.
As the computing requirement of PIMC scales as a moderate power, typically $N^3$, of the system size,
and no attempt has yet been made to parallelize the code,
we expect we can routinely simulate dozens of particles using the method we elaborated.

\section{Conclusion and Outlook}
\label{conclusion}

We have explored the feasibility of the path-integral Monte Carlo simulation of systems
that do not obey time-reversal symmetry under periodic boundary conditions.
Technically, this requires the use of the single-particle thermal density matrix
that is appropriate for the boundary conditions in the presence of a magnetic field.
We have derived several equivalent closed-form expressions for this purpose.
The multi-slice sampling algorithm was modified for the case in which the
weight of a path is determined by the magnitude of the density matrix, which does
not obey a convolution property.
We have illustrated the use of these techniques in the simulation of two-dimensional Yukawa systems,
where time-reversal symmetry is broken by the Coriolis-force,
as commonly done in experiments on cold atomic systems.
We have shown that in spite of the crudeness of the phase-fixing we used, the
interaction-driven transition between a crystalline phase and a correlated liquid
can be captured qualitatively by a PIMC simulation.
A comprehensive quantitative study of this system is delegated to future work.
Eventually, fermions that interact by the Coulomb potential are of more fundamental interest.
For such systems the primitive approximation to the action is clearly not an adequate starting point.
More sophisticated approximations exist, but in their current form they rely upon the
consequences of time-reversal invariance.
The development of suitable approximations for the non-time-reversal-invariant case is underway and is
delegated to future publications.

\begin{acknowledgments}
This research was supported by the National Research Development and Innovation Office of Hungary
within the Quantum Technology National Excellence Program  (Project No.\ 2017-1.2.1-NKP-2017-00001),
and by the Hungarian Scientific Research Funds No.\ K105149.
We are grateful to the HPC facility at the Budapest University of Technology and Economics.
We thank P\'eter L\'evay and Bal\'azs Het\'enyi for useful discussions.
C.\ T.\ was supported by the Hungarian Academy of Sciences.
H.\ G.\ T.\ acknowledges support from the ``Quantum Computing and Quantum Technologies'' PhD School of
the University of Basel.
\end{acknowledgments}

\appendix

\section{The derivation of the single-particle density matrix}

\label{app:dm}

\subsection{Single-particle states on the torus}

In the gauge $\mathbf A = -By\mathbf{\hat x}$ the states in the
\textit{lowest} Landau level assume the form \cite{Haldane85}
\begin{equation}
\label{lowest}
\psi_0(z)=f(z)e^{-\frac{y^2}{2\ell^2}},
\end{equation}
where $f(z)$ is a holomorphic function.
We seek the holomorphic part $f(z)$ of the lowest Landau level eigenstates in terms of
Jacobi elliptic functions, see Eq.~(\ref{eq:theta}).
The twisted boundary conditions we impose in Eq.~(\ref{eq:twisted}) yield $N_\phi$ distinct states \cite{Levay95,Read96},
\begin{equation}
\psi_{0m}(z)=\frac{1}{\sqrt{\ell L_1\sqrt\pi}}\vartheta\begin{bmatrix} a_m \\ b_m\end{bmatrix}
\left(\frac{\pi N_\phi z}{L_1}\Big|N_\phi\tau\right)e^{-\frac{y^2}{2\ell^2}},
\end{equation}
$m=0,1,\dots,(N_\phi-1)$, and $a_m,b_m$ defined in Eq.~(\ref{eq:abdef}).
We note that $\psi_{0m}(z)$, together with its higher Landau level descendants that follow later,
is normalized for the magnetic unit cell,
\begin{multline}
\int_0^{L_2\sin\theta}dy\int_{y\cot\theta}^{y\cot\theta+L_1}dx \psi^\ast_{n'm'}(x+iy)\psi_{nm}(x+iy)=\\
=\delta_{nn'}\delta_{mm'}.
\end{multline}
This particular basis corresponds to a string arrangement \cite{Haldane85} of zeros
of the holomorphic function $f(z)$ in the principal domain.

The orbitals in higher Landau levels are obtained by the application of the Landau level ladder operators,
\begin{equation}
\psi_{nm}(z)=\frac{(a^\dag)^n}{\sqrt{n!}}\psi_{0m}(z),
\end{equation}
where
\begin{equation}
\hat a^\dag = i\ell\sqrt2\left(\partial_z-iA_z\right),
\end{equation}
with $\partial_z=\frac{1}{2}(\partial_x-i\partial_y)$ and $A_z=\frac{1}{2}(A_x-iA_y)$.
In our particular gauge, 
$\hat a^\dag = \frac{\ell}{\sqrt2}\left(i\partial_x+\partial_y-\frac{y}{\ell^2}\right)$.
The degeneracy of each Landau level is $N_\phi$.
Straightforward algebra yields
\begin{multline}
\label{orbitals}
\psi_{nm}(z)=\frac{(-1)^n}{\sqrt{2^nn!\ell L_1\sqrt\pi}}
\sum_{p=-\infty}^\infty
H_n\left(\frac{y+C_{p,m}}{\ell}\right)\\
\times\exp\left(i\pi\tau N_\phi(p+a_m)^2 + 2\pi i(p+a_m)b_m\right)\\
\times\exp\left(\frac{C_{p,m}^2}{2\ell^2}+\frac{iC_{p,m}x}{\ell^2}-\frac{(y+C_{p,m})^2}{2\ell^2}\right),
\end{multline}
where $C_{p,m}=\frac{2\pi N_\phi\ell^2}{L_1}(p+a_m)=L_1(p+a_m)\Im\tau$.

\subsection{The thermal density matrix}

If we substitute Eq.~(\ref{orbitals}) in the definition of the density matrix, Eq.~(\ref{eq:dm}),
the summation over $n$ can be performed by Mehler's formula, and we get
\begin{multline}
\rho^\text{PBC}(\mathbf r,\mathbf r';\beta) =
\frac{\sqrt u}{\ell L_1\sqrt\pi\sqrt{1-u^2}}\sum_{m=0}^{N_\phi-1}\sum_{p,p'=-\infty}^\infty\\
\times\exp\left(i\pi\tau N_\phi(p+a_m)^2-i\pi\tau^\ast N_\phi(p'+a_m)^2 +\right.\\
\left.+ 2\pi i(p+a_m)b_m - 2\pi i(p'+a_m)b_m+\right.\\
\left.+\frac{C_{p,m}^2}{2\ell^2}+\frac{iC_{p,m}x}{\ell^2}+
\frac{C_{p',m}^2}{2\ell^2}-\frac{iC_{p',m}x'}{\ell^2}\right.\\
\left.-\frac{1}{2\ell^2}\frac{1+u^2}{1-u^2}\left((y+C_{p,m})^2+(y'+C_{p',m})^2\right) +\right.\\
\left.+\frac{2u(y+C_{p,m})(y'+C_{p',m})}{(1-u^2)\ell^2}\right).
\end{multline}
Introducing new summation variables $n_1=p+p'$ and $n_2=p-p'$, double-counting is
avoided if $n_1,n_2$ are either both even or both odd.
This decouples the summation variables in all terms except for a factor of
$\exp(i\pi N_\phi n_1n_2\Re\tau)$.
This can be omitted if Eq.~(\ref{eq:restriction}) holds.
As $L_1/N_\phi$ is the separation of the guiding centers of orbitals in the $\mathbf L_1$ direction,
this condition simply means that a translation by $\mathbf L_2$ should be compatible with these
guiding center positions.
By simple algebra and the application of $\vartheta$ functions in Eq.~(\ref{eq:theta}) we obtain
\begin{multline}
\label{firstform}
\rho^\text{PBC}(\mathbf r,\mathbf r';\beta) =
\frac{\sqrt u}{\ell L_1\sqrt\pi\sqrt{1-u^2}}\\
\times\exp\left(
-\frac{1}{2\ell^2}\frac{1+u^2}{1-u^2}(y^2+{y'}^2)+\frac{2u}{1-u^2}\frac{yy'}{\ell^2}\right)\\
\times\sum_{m=0}^{N_\phi-1}\left\{
\vartheta\begin{bmatrix} a_m \\ 0 \end{bmatrix}(z'_1|\tau'_1)
\vartheta\begin{bmatrix} 0 \\ 2b_m' \end{bmatrix}(z_2|\tau_2)+\right.\\
\left.+(-1)^k
\vartheta\begin{bmatrix} a_m+\frac{1}{2} \\ 0 \end{bmatrix}(z'_1|\tau'_1)
\vartheta\begin{bmatrix} \frac{1}{2} \\ 2b_m' \end{bmatrix}(z_2|\tau_2)
\right\},
\end{multline}
where we have used the definitions in Eq.~(\ref{eq:zdef}), and
\begin{equation}
\begin{split}
\tau'_1&=i\pi\left(\frac{2\ell N_\phi}{L_1}\right)^2\frac{1-u}{1+u},\\
z'_1&=\frac{N_\phi\pi}{L_1}\left(x-x' + i(y+y')\frac{1-u}{1+u}\right).
\end{split}
\end{equation}

The density matrix in Eq.~(\ref{firstform}) can be cast in a different form by the application of a modular
transformation $\tau'_1\to\tau_1=-\frac{1}{\tau'_1}$,
$z'_1\to z_1=\frac{z'_1}{\tau'_1}$ in the corresponding $\vartheta$ functions.
The result is Eq.~(\ref{eq:second}).
The structure of Eq.~(\ref{eq:second}) is more transparent perhaps because the $x$- and $y$-components of
the difference vector $\mathbf r-\mathbf r'$ and the center-of-mass vector $\frac{\mathbf r+\mathbf r'}{2}$ appear
on the same footing in the $\vartheta$ functions.

\section{Computational considerations}
\label{app:comput}

While our first formula for the thermal density matrix, Eq.~(\ref{firstform}),
and the one we obtain by a modular transformation, Eq.~(\ref{eq:second}),
are mathematically equivalent, they do differ from a computational point of view.
As each $\vartheta$ function is computed as a sum of Gaussians with subsequently shifted arguments,
it is essential that those Gaussians should be narrow.
This is ensured if the parameters ($\tau_1$, $\tau'_1$, $\tau_2$) of those $\vartheta$'s have a large magnitude.
Notice that $\tau_1$ and $\tau_2$ are pure imaginary, and
\begin{equation}
\begin{split}
\lim_{\beta\to\infty}|\tau'_1|=\lim_{\beta\to\infty}|\tau_2|=\frac{2N_\phi L_2\sin\theta}{L_1},\\
\lim_{\beta\to\infty}|\tau_1|=\frac{L_1}{2 N_\phi L_2\sin\theta},\\
\lim_{\beta\to0}|\tau_1|=\lim_{\beta\to0}|\tau_2|=\infty,\\
\lim_{\beta\to0}|\tau'_1|=0.
\end{split}
\end{equation}
Hence it is advantageous to use Eq.~(\ref{firstform}) for large $\beta$
and Eq.~(\ref{eq:second}) for small $\beta$.
Spelling out the summations implicit in the Jacobi $\vartheta$ functions,
\begin{widetext}
\begin{equation}
\label{compform}
\rho^\text{PBC}(\mathbf r,\mathbf r';\beta) =
\frac{1}{\ell L_1\sqrt\pi}\sqrt{\frac{u}{1-u^2}}\sum_{m=0}^{N_\phi-1}
\left\{
\sum_{n_1=-\infty}^\infty A^{(')}_{0mn_1}\sum_{n_2=-\infty}^\infty B_{0mn_2}^{(')}+
(-1)^k\sum_{n_1=-\infty}^\infty A^{(')}_{\frac{1}{2}mn_1}\sum_{n_2=-\infty}^\infty B^{(')}_{\frac{1}{2}mn_2}
\right\},
\end{equation}
where
\begin{gather}
A_{dmn_1}=\exp\left\{
i\pi\tau'_1\left(n_1+a_m+d+\frac{y+y'}{2L_2\sin\theta}\right)^2 +
 2\pi i N_\phi\left(n_1+a_m+d\right)\frac{x-x'}{L_1}
\right\}\label{largebeta}\\
A'_{dmn_1}=\sqrt\frac{i}{\tau'_1}
\exp\left\{
\frac{i(x'-x)(y+y')}{2\ell^2}+
\frac{\pi}{i\tau'_1}
\left(n_1+N_\phi\frac{x'-x}{L_1}\right)^2 + 2\pi i n_1\left(\frac{y+y'}{2L_2\sin\theta} + a_m + d\right)
\right\},\label{smallbeta}
\end{gather}
and
\begin{gather}
B_{dmn_2}=\exp\left\{
i\pi\tau_2\left(n_2+d+\frac{y-y'}{2L_2\sin\theta}\right)^2 +
2\pi i\left(n_2+d\right)\left(N_\phi\frac{x-x'}{L_1}+2b_m'\right)
\right\},\label{largebeta2}\\
B'_{dmn_2}=\label{smallbeta2}
\sqrt\frac{i}{\tau_2}
\exp\left\{\frac{i(y'-y)(x+x')}{2\ell^2}+\frac{2 \pi ib_m'(y'-y)}{L_2\sin\theta}+
\frac{\pi}{i\tau_2}
\left(n_2-N_\phi\frac{x+x'}{L_1}-2b_m'\right)^2 + 2\pi i n_2\left(\frac{y-y'}{2L_2\sin\theta} + d\right)\right\},
\end{gather}
\end{widetext}
Here, the $A'$, $B'$ terms come from Eq.~(\ref{eq:second}) and the unprimed ones are from Eq.~(\ref{firstform}).
Notice that $A'_{dmn_1}\neq A_{dmn_1}$ and $B'_{dmn_1}\neq B_{dmn_1}$,
the primed and unprimed expressions are interchangeable only within the summation over $n_1$ and $n_2$, respectively.
We have found it convenient to use Eq.~(\ref{largebeta}) in the low-temperature range
$\tanh\left(\frac{\beta\hbar\omega_c}{2}\right)>\frac{L_1}{2N_\phi L_2\sin\theta}$,
and Eq.~(\ref{smallbeta}) otherwise (high temperature).
For the other term, $B_{dmn_2}$ in Eq.~(\ref{largebeta2})
is almost always preferable to $B'_{dmn_2}$ in Eq.~(\ref{smallbeta2}),
except if $N_\phi$ and $\theta$ are small and $\beta$ large.
Using Eq.~(\ref{eq:abdef}), $B_{dmn_2}$ is independent of $m$ iff $\Re\tau$ is an integer, i.e.,
\begin{equation}
\label{stricteq}
k'=\frac{k}{N_\phi}
\end{equation}
is an integer.
Notice that this condition is stricter than Eq.~(\ref{eq:restriction}).
(Both conditions hold trivially for a rectangular torus.)
Then, using $A'_{mn_1d}$ in Eq.~(\ref{smallbeta}) and $B_{mn_2d}$ in Eq.~(\ref{largebeta2}),
the summation over $m$ can be performed.
If, furthermore, $N_\phi$ is even, an extremely compact formula is obtained:
\begin{multline}
\rho^\text{PBC}(\mathbf r,\mathbf r';\beta) =
\frac{1}{2\pi\ell^2}\frac{\sqrt u}{1-u}
\exp\left(\frac{i(x'-x)(y+y')}{2\ell^2}\right)\\
\times\sum_{n_1=-\infty}^\infty
\exp\left(-\frac{1+u}{1-u}\frac{1}{4\ell^2}\left(x-x'-n_1L_1\right)^2+\right.\\
\left.+i\pi n_1\left(N_\phi\frac{y+y'}{L_2\sin\theta}+\frac{\phi_1}{\pi}\right)\right)\\
\times\sum_{n_2=-\infty}^\infty
\exp\left(-\frac{1+u}{1-u}\frac{1}{4\ell^2}\left(y-y'+n_2L_2\sin\theta\right)^2+\right.\\
\left.+i\pi n_2\left(N_\phi\frac{x+x'}{L_1}-\frac{\phi_2-k'\phi_1}{\pi}
\right)
\right).
\label{compact}
\end{multline}
Notice that Eq.~(\ref{compact}) amounts to obtaining the density matrix for twisted periodic boundary conditions
from the corresponding object for the infinite plain [Eq.~(\ref{eq:openbc})] as the sum
\begin{equation}
\sum_{n_1,n_2=-\infty}^\infty
e^{-in_1\phi_1-in_2\phi_2}t_{\mathbf r}(n_1\mathbf L_1+n_2\mathbf L_2)
\rho^\text{open}(\mathbf r,\mathbf r';\beta).
\end{equation}
However, the two infinite summations in this formula do not decouple unless the condition in
Eq.~(\ref{stricteq}) holds and $N_\phi$ is even.

\section{Phase fixing}
\label{app:pfaction}

As phase fixing for PIMC has already been described in the literature \cite{Akkineni08},
we just review the relevant formulas for completeness.
The thermal density matrix satisfies Bloch's equation
\begin{equation}
\label{eq:bloch}
\frac{\partial}{\partial\beta}\rho(R,R';\beta) = \mathcal H \rho(R,R';\beta),
\end{equation}
where
\begin{equation}
\mathcal H=\sum_{i=1}^N\lambda\left(\nabla_i-\frac{e}{\hbar}\mathbf A(\mathbf r_i)\right)^2 + V(R)
\end{equation}
is the Hamiltonian that acts on the unprimed coordinates, and $\lambda=\frac{\hbar^2}{2m}$.
We let $\nabla\equiv (\nabla_1,\dots,\nabla_N)$ and
$A(R)\equiv (\mathbf A(\mathbf r_1),\dots,\mathbf A(\mathbf r_N))$.
Separating the magnitude and the phase of the density matrix as
\begin{equation}
\rho(R,R';\beta) =|\rho(R,R';\beta)|e^{i\varphi(R,R';\beta)},
\end{equation}
Eq.~(\ref{eq:bloch}) maps to two coupled partial differential equations
\begin{eqnarray}
\label{eq:magnitude}
\frac{\partial|\rho|}{\partial\beta} &=& \lambda\nabla^2|\rho|
-\left[V + \lambda\left(\nabla\varphi - \frac{e}{\hbar}A\right)^2\right]|\rho|,\\
\frac{\partial\varphi}{\partial\beta} &=& \lambda\left(
\nabla^2\varphi + 2\frac{\nabla|\rho|\cdot\nabla\varphi}{|\rho|}
-2\frac{e}{\hbar}\frac{A\cdot\nabla|\rho|}{|\rho|}
-\frac{e}{\hbar}\nabla\cdot A\right),\nonumber
\end{eqnarray}
where we have suppressed the arguments $(R,R';\beta)$ for $\rho$ and $\varphi$, and $(R)$ for $V$ and $A$, respectively.
Consider some variational many-body density matrix $\rho_T(R,R';\beta) =|\rho_T(R,R';\beta)|e^{i\varphi_T(R,R';\beta)}$.
We seek the density matrix $\rho(R,R';\beta)$ under the assumption that $\varphi(R,R';\beta)=\varphi_T(R,R';\beta)$,
i.e., with its phase fixed.
Then Eq.~(\ref{eq:magnitude}) is formally equivalent to a Bloch equation for $|\rho(R,R';\beta)|$
with effective potential ($R'$ is fixed)
\begin{equation}
V_\text{eff}(R) = V(R) + \lambda\left(\nabla\varphi_T(R,R';\beta) - \frac{e}{\hbar}A(R)\right)^2.
\end{equation}
Thus PIMC with phase fixing samples paths with real and nonnegative weight, using a
fixed-phase dependent effective interaction.

If we know $\varphi_T(R_m,R_{m-1};\beta)$ and its gradient $\nabla_{R_m}\varphi_T(R_m,R_{m-1};\beta)$,
we can apply the following approximation.
The gradient is decomposed into components parallel and perpendicular to the semiclassical path between $(R_{m-1},0)$ and  $(R_{m},\tau)$:
\begin{equation}
\label{eq:pp}
\begin{split}
G^\parallel(R)&=\nabla\varphi_T(R)\cdot\frac{R_{m}-R_{m-1}}{|R_{m}-R_{m-1}|},\\
G^\perp(R)&=\sqrt{|\nabla\varphi_T(R)|^2 - (G^\parallel(R))^2}.
\end{split}
\end{equation}
The perpendicular component is taken into account by the primitive action.
On the other hand, the evolution of the phase is approximated by a cubic polynomial on the semiclassical trajectory, and
the contribution of the parallel component of the gradient of $\varphi_T$ is integrated
on this trajectory as in the semiclassical approximation to the action.
Technically, we assume the following quantities are known:
\begin{equation}
\begin{split}
\varphi_1&=\lim_{\tau^\ast\to0}\lim_{R\to R_{m-1}}\varphi_T(R,R_{m-1};\tau^\ast)=0,\\
g_1&=\lim_{\tau^\ast\to0}\lim_{R\to R_{m-1}}\nabla_R\varphi_T(R,R_{m-1};\tau^\ast),\\
\varphi_2&=\varphi_T(R_{m},R_{m-1};\tau),\\
g_2&=\nabla_{R_m}\varphi_T(R_{m},R_{m-1};\tau),
\end{split}
\end{equation}
and $g_1^\perp$, $g_1^\parallel$, $g_2^\perp$, $g_2^\parallel$ are magnitudes of the perpendicular
and parallel components of $g_1$ and $g_2$, respectively, in the sense of Eq.~(\ref{eq:pp}).
(If the phase is fixed to a single-particle density matrix, $g_1=-y'\mathbf{\hat x}/\ell^2$ both for open and periodic boundary conditions.
If the phase of the free Fermi or Bose gas is used, cf.\ Eqs.~(\ref{eq:freefermi}-\ref{eq:freebose}),
$g_1=-\sum_iy'_i\mathbf{\hat x_i}/\ell^2$.)

The perpendicular component is taken into account by the primitive action:
\begin{equation}
\label{eq:ufp0}
U_\text{FP,0}(R_{m},R_{m-1};\tau) = \frac{\lambda\tau}{2}\left((g_1^\perp)^2 + (g_2^\perp)^2\right).
\end{equation}
The next contribution is the line integral of $(G^\parallel)^2$ on the straight path between $R_{m-1}$ and
$R_m$, if $\varphi_T$ is approximated by a cubic polynomial on this route.
\begin{multline}
U_\text{FP,1}(R_{m},R_{m-1};\tau) = \frac{\lambda\tau}{15}\left[
2\left((g_1^\parallel)^2 + (g_2^\parallel)^2)\right)-g_1^\parallel g_2^\parallel-\right.\\
\left.-3\frac{(g_1^\parallel +g_2^\parallel)(\varphi_2-\varphi_1)}{\delta R}
+18\frac{(\varphi_2-\varphi_1)^2}{\delta R^2}
\right].
\end{multline}
We proceed in the same way for the dot product of the phase gradient and the vector potential.
$A\cdot G^\perp$ contributes at the end points:
\begin{multline}
U_\text{FP,2}(R_{m},R_{m-1};\tau) = \frac{\lambda}{\ell^2}\sum_{j=1}^2
g_{j}^\perp\\
\times\sqrt{
\sum_{i=1}^N y_{m-1+j,i}^2 - \left(\frac{\sum_{i=1}^N y_{m-1+j,i} (x_{m,i}-x_{m-1,i})}{\delta R}\right)^2},
\end{multline}
and for $A\cdot G^\parallel$ we again use the semiclassical action with the cubic
approximation for $\varphi_T$:
\begin{multline}
U_\text{FP,3}(R_{m},R_{m-1};\tau) = \frac{2\lambda}{\ell^2}\sum_{i=1}^N (x_{m,1}-x_{m-1,i})\\
\times\left[
\left(\frac{c_3\delta R}{4}+\frac{c_2}{3}+\frac{g^\parallel_1}{2\delta R}\right)y_{m-1,i}+\right.\\
\left.+\left(\frac{3c_3\delta R}{4}+\frac{2c_2}{3}+\frac{g^\parallel_1}{2\delta R}\right)y_{m,i}
\right],
\end{multline}
where $c_2=\frac{3(\varphi_2-\varphi_1)}{\delta R^2} - \frac{2g^\parallel_1+g^\parallel_2}{\delta R}$ and
$c_3=\frac{g^\parallel_1+g^\parallel_2}{\delta R^2} - \frac{2(\varphi_2-\varphi_1)}{\delta R^3}$.
Finally, the semiclassical contribution of the $A^2$ term is
\begin{multline}
\label{eq:ufp4}
U_\text{FP,4}(R_{m},R_{m-1};\tau) = \frac{\lambda}{3\ell^4}\\
\times\sum_{i=1}^N\left(
y_{m-1,i}^2 + y_{m,i}^2 + y_{m-1,i}y_{m,i}
\right).
\end{multline}
The total contribution is the sum of Eqs.~(\ref{eq:ufp0}) to (\ref{eq:ufp4}).

\bibliography{qmcbiblio}

\begin{thebibliography}{38}
\expandafter\ifx\csname natexlab\endcsname\relax\def\natexlab#1{#1}\fi
\expandafter\ifx\csname bibnamefont\endcsname\relax
  \def\bibnamefont#1{#1}\fi
\expandafter\ifx\csname bibfnamefont\endcsname\relax
  \def\bibfnamefont#1{#1}\fi
\expandafter\ifx\csname citenamefont\endcsname\relax
  \def\citenamefont#1{#1}\fi
\expandafter\ifx\csname url\endcsname\relax
  \def\url#1{\texttt{#1}}\fi
\expandafter\ifx\csname urlprefix\endcsname\relax\def\urlprefix{URL }\fi
\providecommand{\bibinfo}[2]{#2}
\providecommand{\eprint}[2][]{\url{#2}}

\bibitem[{\citenamefont{Ceperley}(1995)}]{Ceperley95}
\bibinfo{author}{\bibfnamefont{D.~M.} \bibnamefont{Ceperley}},
  \bibinfo{journal}{Rev. Mod. Phys.} \textbf{\bibinfo{volume}{67}},
  \bibinfo{pages}{279} (\bibinfo{year}{1995}),
  \urlprefix\url{http://link.aps.org/doi/10.1103/RevModPhys.67.279}.

\bibitem[{\citenamefont{Metropolis et~al.}(1953)\citenamefont{Metropolis,
  Rosenbluth, Rosenbluth, Teller, and Teller}}]{Metropolis53}
\bibinfo{author}{\bibfnamefont{N.}~\bibnamefont{Metropolis}},
  \bibinfo{author}{\bibfnamefont{A.~W.} \bibnamefont{Rosenbluth}},
  \bibinfo{author}{\bibfnamefont{M.~N.} \bibnamefont{Rosenbluth}},
  \bibinfo{author}{\bibfnamefont{A.~H.} \bibnamefont{Teller}},
  \bibnamefont{and} \bibinfo{author}{\bibfnamefont{E.}~\bibnamefont{Teller}},
  \bibinfo{journal}{The Journal of Chemical Physics}
  \textbf{\bibinfo{volume}{21}}, \bibinfo{pages}{1087} (\bibinfo{year}{1953}),
  \eprint{http://dx.doi.org/10.1063/1.1699114},
  \urlprefix\url{http://dx.doi.org/10.1063/1.1699114}.

\bibitem[{\citenamefont{Hastings}(1970)}]{Hastings70}
\bibinfo{author}{\bibfnamefont{W.~K.} \bibnamefont{Hastings}},
  \bibinfo{journal}{Biometrika} \textbf{\bibinfo{volume}{57}},
  \bibinfo{pages}{97} (\bibinfo{year}{1970}),
  \urlprefix\url{http://dx.doi.org/10.1093/biomet/57.1.97}.

\bibitem[{\citenamefont{Ceperley}(1991)}]{Ceperley91}
\bibinfo{author}{\bibfnamefont{D.~M.} \bibnamefont{Ceperley}},
  \bibinfo{journal}{Journal of Statistical Physics}
  \textbf{\bibinfo{volume}{63}}, \bibinfo{pages}{1237} (\bibinfo{year}{1991}),
  ISSN \bibinfo{issn}{1572-9613},
  \urlprefix\url{https://doi.org/10.1007/BF01030009}.

\bibitem[{\citenamefont{Ceperley}(1996)}]{Ceperley96}
\bibinfo{author}{\bibfnamefont{D.~M.} \bibnamefont{Ceperley}}, in
  \emph{\bibinfo{booktitle}{Monte Carlo and molecular dynamics of condensed
  matter systems}}, edited by
  \bibinfo{editor}{\bibfnamefont{K.}~\bibnamefont{Binder}} \bibnamefont{and}
  \bibinfo{editor}{\bibfnamefont{G.}~\bibnamefont{Ciccotti}}
  (\bibinfo{publisher}{Italian Physical Society}, \bibinfo{year}{1996}).

\bibitem[{\citenamefont{Ortiz et~al.}(1993)\citenamefont{Ortiz, Ceperley, and
  Martin}}]{Ortiz93}
\bibinfo{author}{\bibfnamefont{G.}~\bibnamefont{Ortiz}},
  \bibinfo{author}{\bibfnamefont{D.~M.} \bibnamefont{Ceperley}},
  \bibnamefont{and} \bibinfo{author}{\bibfnamefont{R.~M.}
  \bibnamefont{Martin}}, \bibinfo{journal}{Phys. Rev. Lett.}
  \textbf{\bibinfo{volume}{71}}, \bibinfo{pages}{2777} (\bibinfo{year}{1993}),
  \urlprefix\url{http://link.aps.org/doi/10.1103/PhysRevLett.71.2777}.

\bibitem[{\citenamefont{Bolton}(1996)}]{Bolton96}
\bibinfo{author}{\bibfnamefont{F.}~\bibnamefont{Bolton}},
  \bibinfo{journal}{Phys. Rev. B} \textbf{\bibinfo{volume}{54}},
  \bibinfo{pages}{4780} (\bibinfo{year}{1996}),
  \urlprefix\url{https://link.aps.org/doi/10.1103/PhysRevB.54.4780}.

\bibitem[{\citenamefont{Foulkes et~al.}(2001)\citenamefont{Foulkes, Mitas,
  Needs, and Rajagopal}}]{Foulkes01}
\bibinfo{author}{\bibfnamefont{W.~M.~C.} \bibnamefont{Foulkes}},
  \bibinfo{author}{\bibfnamefont{L.}~\bibnamefont{Mitas}},
  \bibinfo{author}{\bibfnamefont{R.~J.} \bibnamefont{Needs}}, \bibnamefont{and}
  \bibinfo{author}{\bibfnamefont{G.}~\bibnamefont{Rajagopal}},
  \bibinfo{journal}{Rev. Mod. Phys.} \textbf{\bibinfo{volume}{73}},
  \bibinfo{pages}{33} (\bibinfo{year}{2001}),
  \urlprefix\url{http://link.aps.org/doi/10.1103/RevModPhys.73.33}.

\bibitem[{\citenamefont{Akkineni}(2008)}]{Akkineni08}
\bibinfo{author}{\bibfnamefont{V.}~\bibnamefont{Akkineni}},
  \bibinfo{type}{dissertation}, \bibinfo{school}{University of Illinois at
  Urbana-Champaign} (\bibinfo{year}{2008}),
  \urlprefix\url{http://hdl.handle.net/2142/80567}.

\bibitem[{\citenamefont{Melik-Alaverdian and
  Bonesteel}(1995)}]{Melik-Alaverdian95}
\bibinfo{author}{\bibfnamefont{V.}~\bibnamefont{Melik-Alaverdian}}
  \bibnamefont{and} \bibinfo{author}{\bibfnamefont{N.~E.}
  \bibnamefont{Bonesteel}}, \bibinfo{journal}{Phys. Rev. B}
  \textbf{\bibinfo{volume}{52}}, \bibinfo{pages}{R17032}
  (\bibinfo{year}{1995}),
  \urlprefix\url{http://link.aps.org/doi/10.1103/PhysRevB.52.R17032}.

\bibitem[{\citenamefont{Melik-Alaverdian
  et~al.}(1997)\citenamefont{Melik-Alaverdian, Bonesteel, and
  Ortiz}}]{Melik-Alaverdian97}
\bibinfo{author}{\bibfnamefont{V.}~\bibnamefont{Melik-Alaverdian}},
  \bibinfo{author}{\bibfnamefont{N.~E.} \bibnamefont{Bonesteel}},
  \bibnamefont{and} \bibinfo{author}{\bibfnamefont{G.}~\bibnamefont{Ortiz}},
  \bibinfo{journal}{Phys. Rev. Lett.} \textbf{\bibinfo{volume}{79}},
  \bibinfo{pages}{5286} (\bibinfo{year}{1997}),
  \urlprefix\url{http://link.aps.org/doi/10.1103/PhysRevLett.79.5286}.

\bibitem[{\citenamefont{Lin et~al.}(2001)\citenamefont{Lin, Zong, and
  Ceperley}}]{Lin01}
\bibinfo{author}{\bibfnamefont{C.}~\bibnamefont{Lin}},
  \bibinfo{author}{\bibfnamefont{F.~H.} \bibnamefont{Zong}}, \bibnamefont{and}
  \bibinfo{author}{\bibfnamefont{D.~M.} \bibnamefont{Ceperley}},
  \bibinfo{journal}{Phys. Rev. E} \textbf{\bibinfo{volume}{64}},
  \bibinfo{pages}{016702} (\bibinfo{year}{2001}),
  \urlprefix\url{http://link.aps.org/doi/10.1103/PhysRevE.64.016702}.

\bibitem[{\citenamefont{Shao et~al.}(2015)\citenamefont{Shao, Kim, Haldane, and
  Rezayi}}]{Shao15}
\bibinfo{author}{\bibfnamefont{J.}~\bibnamefont{Shao}},
  \bibinfo{author}{\bibfnamefont{E.-A.} \bibnamefont{Kim}},
  \bibinfo{author}{\bibfnamefont{F.~D.~M.} \bibnamefont{Haldane}},
  \bibnamefont{and} \bibinfo{author}{\bibfnamefont{E.~H.}
  \bibnamefont{Rezayi}}, \bibinfo{journal}{Phys. Rev. Lett.}
  \textbf{\bibinfo{volume}{114}}, \bibinfo{pages}{206402}
  (\bibinfo{year}{2015}),
  \urlprefix\url{http://link.aps.org/doi/10.1103/PhysRevLett.114.206402}.

\bibitem[{\citenamefont{Pollock and Ceperley}(1984)}]{Pollock84}
\bibinfo{author}{\bibfnamefont{E.~L.} \bibnamefont{Pollock}} \bibnamefont{and}
  \bibinfo{author}{\bibfnamefont{D.~M.} \bibnamefont{Ceperley}},
  \bibinfo{journal}{Phys. Rev. B} \textbf{\bibinfo{volume}{30}},
  \bibinfo{pages}{2555} (\bibinfo{year}{1984}),
  \urlprefix\url{https://link.aps.org/doi/10.1103/PhysRevB.30.2555}.

\bibitem[{\citenamefont{Nelson and Seung}(1989)}]{Nelson89}
\bibinfo{author}{\bibfnamefont{D.~R.} \bibnamefont{Nelson}} \bibnamefont{and}
  \bibinfo{author}{\bibfnamefont{H.~S.} \bibnamefont{Seung}},
  \bibinfo{journal}{Phys. Rev. B} \textbf{\bibinfo{volume}{39}},
  \bibinfo{pages}{9153} (\bibinfo{year}{1989}),
  \urlprefix\url{https://link.aps.org/doi/10.1103/PhysRevB.39.9153}.

\bibitem[{\citenamefont{Magro and Ceperley}(1993)}]{Magro93}
\bibinfo{author}{\bibfnamefont{W.~R.} \bibnamefont{Magro}} \bibnamefont{and}
  \bibinfo{author}{\bibfnamefont{D.~M.} \bibnamefont{Ceperley}},
  \bibinfo{journal}{Phys. Rev. B} \textbf{\bibinfo{volume}{48}},
  \bibinfo{pages}{411} (\bibinfo{year}{1993}),
  \urlprefix\url{https://link.aps.org/doi/10.1103/PhysRevB.48.411}.

\bibitem[{\citenamefont{Nordborg and Blatter}(1997)}]{Nordborg97}
\bibinfo{author}{\bibfnamefont{H.}~\bibnamefont{Nordborg}} \bibnamefont{and}
  \bibinfo{author}{\bibfnamefont{G.}~\bibnamefont{Blatter}},
  \bibinfo{journal}{Phys. Rev. Lett.} \textbf{\bibinfo{volume}{79}},
  \bibinfo{pages}{1925} (\bibinfo{year}{1997}),
  \urlprefix\url{https://link.aps.org/doi/10.1103/PhysRevLett.79.1925}.

\bibitem[{\citenamefont{Petrov et~al.}(2007)\citenamefont{Petrov,
  Astrakharchik, Papoular, Salomon, and Shlyapnikov}}]{Petrov07}
\bibinfo{author}{\bibfnamefont{D.~S.} \bibnamefont{Petrov}},
  \bibinfo{author}{\bibfnamefont{G.~E.} \bibnamefont{Astrakharchik}},
  \bibinfo{author}{\bibfnamefont{D.~J.} \bibnamefont{Papoular}},
  \bibinfo{author}{\bibfnamefont{C.}~\bibnamefont{Salomon}}, \bibnamefont{and}
  \bibinfo{author}{\bibfnamefont{G.~V.} \bibnamefont{Shlyapnikov}},
  \bibinfo{journal}{Phys. Rev. Lett.} \textbf{\bibinfo{volume}{99}},
  \bibinfo{pages}{130407} (\bibinfo{year}{2007}),
  \urlprefix\url{https://link.aps.org/doi/10.1103/PhysRevLett.99.130407}.

\bibitem[{\citenamefont{Madison et~al.}(2000)\citenamefont{Madison, Chevy,
  Wohlleben, and Dalibard}}]{Madison00}
\bibinfo{author}{\bibfnamefont{K.~W.} \bibnamefont{Madison}},
  \bibinfo{author}{\bibfnamefont{F.}~\bibnamefont{Chevy}},
  \bibinfo{author}{\bibfnamefont{W.}~\bibnamefont{Wohlleben}},
  \bibnamefont{and} \bibinfo{author}{\bibfnamefont{J.}~\bibnamefont{Dalibard}},
  \bibinfo{journal}{Phys. Rev. Lett.} \textbf{\bibinfo{volume}{84}},
  \bibinfo{pages}{806} (\bibinfo{year}{2000}).

\bibitem[{\citenamefont{Abo-Shaeer et~al.}(2001)\citenamefont{Abo-Shaeer,
  Raman, Vogels, and Ketterle}}]{AboShaeer01}
\bibinfo{author}{\bibfnamefont{J.~R.} \bibnamefont{Abo-Shaeer}},
  \bibinfo{author}{\bibfnamefont{C.}~\bibnamefont{Raman}},
  \bibinfo{author}{\bibfnamefont{J.~M.} \bibnamefont{Vogels}},
  \bibnamefont{and} \bibinfo{author}{\bibfnamefont{W.}~\bibnamefont{Ketterle}},
  \bibinfo{journal}{Science} \textbf{\bibinfo{volume}{292}},
  \bibinfo{pages}{476} (\bibinfo{year}{2001}).

\bibitem[{\citenamefont{Hodby et~al.}(2001)\citenamefont{Hodby, Hechenblaikner,
  Hopkins, Marag\`o, and Foot}}]{Hodby01}
\bibinfo{author}{\bibfnamefont{E.}~\bibnamefont{Hodby}},
  \bibinfo{author}{\bibfnamefont{G.}~\bibnamefont{Hechenblaikner}},
  \bibinfo{author}{\bibfnamefont{S.~A.} \bibnamefont{Hopkins}},
  \bibinfo{author}{\bibfnamefont{O.~M.} \bibnamefont{Marag\`o}},
  \bibnamefont{and} \bibinfo{author}{\bibfnamefont{C.~J.} \bibnamefont{Foot}},
  \bibinfo{journal}{Phys. Rev. Lett.} \textbf{\bibinfo{volume}{88}},
  \bibinfo{pages}{010405} (\bibinfo{year}{2001}),
  \urlprefix\url{https://link.aps.org/doi/10.1103/PhysRevLett.88.010405}.

\bibitem[{\citenamefont{Haljan et~al.}(2001)\citenamefont{Haljan, Coddington,
  Engels, and Cornell}}]{Haljan01}
\bibinfo{author}{\bibfnamefont{P.~C.} \bibnamefont{Haljan}},
  \bibinfo{author}{\bibfnamefont{I.}~\bibnamefont{Coddington}},
  \bibinfo{author}{\bibfnamefont{P.}~\bibnamefont{Engels}}, \bibnamefont{and}
  \bibinfo{author}{\bibfnamefont{E.~A.} \bibnamefont{Cornell}},
  \bibinfo{journal}{Phys. Rev. Lett.} \textbf{\bibinfo{volume}{87}},
  \bibinfo{pages}{210403} (\bibinfo{year}{2001}).

\bibitem[{\citenamefont{Ceperley}(1992)}]{Ceperley92}
\bibinfo{author}{\bibfnamefont{D.~M.} \bibnamefont{Ceperley}},
  \bibinfo{journal}{Phys. Rev. Lett.} \textbf{\bibinfo{volume}{69}},
  \bibinfo{pages}{331} (\bibinfo{year}{1992}),
  \urlprefix\url{https://link.aps.org/doi/10.1103/PhysRevLett.69.331}.

\bibitem[{\citenamefont{Pierleoni et~al.}(1994)\citenamefont{Pierleoni,
  Ceperley, Bernu, and Magro}}]{Pierleoni94}
\bibinfo{author}{\bibfnamefont{C.}~\bibnamefont{Pierleoni}},
  \bibinfo{author}{\bibfnamefont{D.~M.} \bibnamefont{Ceperley}},
  \bibinfo{author}{\bibfnamefont{B.}~\bibnamefont{Bernu}}, \bibnamefont{and}
  \bibinfo{author}{\bibfnamefont{W.~R.} \bibnamefont{Magro}},
  \bibinfo{journal}{Phys. Rev. Lett.} \textbf{\bibinfo{volume}{73}},
  \bibinfo{pages}{2145} (\bibinfo{year}{1994}),
  \urlprefix\url{https://link.aps.org/doi/10.1103/PhysRevLett.73.2145}.

\bibitem[{\citenamefont{Magro et~al.}(1996)\citenamefont{Magro, Ceperley,
  Pierleoni, and Bernu}}]{Magro96}
\bibinfo{author}{\bibfnamefont{W.~R.} \bibnamefont{Magro}},
  \bibinfo{author}{\bibfnamefont{D.~M.} \bibnamefont{Ceperley}},
  \bibinfo{author}{\bibfnamefont{C.}~\bibnamefont{Pierleoni}},
  \bibnamefont{and} \bibinfo{author}{\bibfnamefont{B.}~\bibnamefont{Bernu}},
  \bibinfo{journal}{Phys. Rev. Lett.} \textbf{\bibinfo{volume}{76}},
  \bibinfo{pages}{1240} (\bibinfo{year}{1996}),
  \urlprefix\url{https://link.aps.org/doi/10.1103/PhysRevLett.76.1240}.

\bibitem[{\citenamefont{Zak}(1964)}]{Zak64}
\bibinfo{author}{\bibfnamefont{J.}~\bibnamefont{Zak}}, \bibinfo{journal}{Phys.
  Rev.} \textbf{\bibinfo{volume}{134}}, \bibinfo{pages}{A1602}
  (\bibinfo{year}{1964}),
  \urlprefix\url{http://link.aps.org/doi/10.1103/PhysRev.134.A1602}.

\bibitem[{com({\natexlab{a}})}]{comment3}
\bibinfo{note}{Just like its time-reversal symmetric counterpart,
  $\rho^\text{PBC}(\mathbf r,\mathbf r';\beta)$ describes diffusive motion in
  imaginary time (inverse temperature), but it also has a gauge-dependent phase
  factor. Compared to the case without the external magnetic field, the
  diffusion process described by Eq.~(\ref{eq:openbc}) is slower; for large
  imaginary time $\beta$ the width of the Gaussian $|\rho^\text{open}(\mathbf
  r,\mathbf r';\beta)|$ tends to a finite value determined by the magnetic
  length. In the other limit, $\beta\hbar\omega_c\ll1$ this Gaussian grows with
  the same rate as in the absence of the magnetic field.}

\bibitem[{\citenamefont{Mumford}(1987)}]{Mumford87}
\bibinfo{author}{\bibfnamefont{D.}~\bibnamefont{Mumford}},
  \emph{\bibinfo{title}{Tata Lectures on Theta I}}, Modern Birkh\"auser
  Classics (\bibinfo{publisher}{Springer}, \bibinfo{year}{1987}), ISBN
  \bibinfo{isbn}{9780817645779},
  \urlprefix\url{http://www.springer.com/us/book/9780817645724}.

\bibitem[{com({\natexlab{b}})}]{comment1}
\bibinfo{note}{Traditionally defined Jacobi elliptic functions correspond to
  $\vartheta\begin{bmatrix} \frac{1}{2} \\ \frac{1}{2} \end{bmatrix}(z|\tau) =
  -\vartheta_1(z|\tau)$, $\vartheta\begin{bmatrix} \frac{1}{2} \\ 0
  \end{bmatrix}(z|\tau) = \vartheta_2(z|\tau)$, $\vartheta\begin{bmatrix} 0 \\
  0 \end{bmatrix}(z|\tau)= \vartheta_3(z|\tau)$, and $\vartheta\begin{bmatrix}
  0 \\ \frac{1}{2} \end{bmatrix}(z|\tau) = \vartheta_4(z|\tau)$. Notice that
  sometimes a factor $\pi$ is factored out of the argument, i.e.,
  $\vartheta\begin{bmatrix} a \\ b
  \end{bmatrix}\left(\frac{z}{\pi}\big|\tau\right)$ is defined.}

\bibitem[{\citenamefont{Aharonov and Bohm}(1959)}]{Aharonov59}
\bibinfo{author}{\bibfnamefont{Y.}~\bibnamefont{Aharonov}} \bibnamefont{and}
  \bibinfo{author}{\bibfnamefont{D.}~\bibnamefont{Bohm}},
  \bibinfo{journal}{Phys. Rev.} \textbf{\bibinfo{volume}{115}},
  \bibinfo{pages}{485} (\bibinfo{year}{1959}),
  \urlprefix\url{http://link.aps.org/doi/10.1103/PhysRev.115.485}.

\bibitem[{\citenamefont{Byers and Yang}(1961)}]{Byers61}
\bibinfo{author}{\bibfnamefont{N.}~\bibnamefont{Byers}} \bibnamefont{and}
  \bibinfo{author}{\bibfnamefont{C.~N.} \bibnamefont{Yang}},
  \bibinfo{journal}{Phys. Rev. Lett.} \textbf{\bibinfo{volume}{7}},
  \bibinfo{pages}{46} (\bibinfo{year}{1961}),
  \urlprefix\url{http://link.aps.org/doi/10.1103/PhysRevLett.7.46}.

\bibitem[{\citenamefont{Cao}(1994)}]{Cao94}
\bibinfo{author}{\bibfnamefont{J.}~\bibnamefont{Cao}}, \bibinfo{journal}{Phys.
  Rev. E} \textbf{\bibinfo{volume}{49}}, \bibinfo{pages}{882}
  (\bibinfo{year}{1994}),
  \urlprefix\url{https://link.aps.org/doi/10.1103/PhysRevE.49.882}.

\bibitem[{\citenamefont{Wilkin et~al.}(1998)\citenamefont{Wilkin, Gunn, and
  Smith}}]{Wilkin98}
\bibinfo{author}{\bibfnamefont{N.~K.} \bibnamefont{Wilkin}},
  \bibinfo{author}{\bibfnamefont{J.~M.~F.} \bibnamefont{Gunn}},
  \bibnamefont{and} \bibinfo{author}{\bibfnamefont{R.~A.} \bibnamefont{Smith}},
  \bibinfo{journal}{Phys. Rev. Lett.} \textbf{\bibinfo{volume}{80}},
  \bibinfo{pages}{2265} (\bibinfo{year}{1998}),
  \urlprefix\url{http://link.aps.org/doi/10.1103/PhysRevLett.80.2265}.

\bibitem[{\citenamefont{Cooper}(2008)}]{Cooper08}
\bibinfo{author}{\bibfnamefont{N.~R.} \bibnamefont{Cooper}},
  \bibinfo{journal}{Advances in Physics} \textbf{\bibinfo{volume}{57}},
  \bibinfo{pages}{539} (\bibinfo{year}{2008}).

\bibitem[{com({\natexlab{c}})}]{comment2}
\bibinfo{note}{Due to the computational complexity of the permanent,
  phase-fixing by the free Bose gas is actually more difficult than by the free
  Fermi gas. In this case study we will restrict our work to relatively small
  systems.}

\bibitem[{\citenamefont{Haldane and Rezayi}(1985)}]{Haldane85}
\bibinfo{author}{\bibfnamefont{F.~D.~M.} \bibnamefont{Haldane}}
  \bibnamefont{and} \bibinfo{author}{\bibfnamefont{E.~H.}
  \bibnamefont{Rezayi}}, \bibinfo{journal}{Phys. Rev. B}
  \textbf{\bibinfo{volume}{31}}, \bibinfo{pages}{2529} (\bibinfo{year}{1985}),
  \urlprefix\url{http://link.aps.org/doi/10.1103/PhysRevB.31.2529}.

\bibitem[{\citenamefont{L\'evay}(1995)}]{Levay95}
\bibinfo{author}{\bibfnamefont{P.}~\bibnamefont{L\'evay}},
  \bibinfo{journal}{Journal of Mathematical Physics}
  \textbf{\bibinfo{volume}{36}}, \bibinfo{pages}{2792} (\bibinfo{year}{1995}),
  \urlprefix\url{http://scitation.aip.org/content/aip/journal/jmp/36/6/10.1063%
/1.531066}.

\bibitem[{\citenamefont{Read and Rezayi}(1996)}]{Read96}
\bibinfo{author}{\bibfnamefont{N.}~\bibnamefont{Read}} \bibnamefont{and}
  \bibinfo{author}{\bibfnamefont{E.}~\bibnamefont{Rezayi}},
  \bibinfo{journal}{Phys. Rev. B} \textbf{\bibinfo{volume}{54}},
  \bibinfo{pages}{16864} (\bibinfo{year}{1996}),
  \urlprefix\url{http://link.aps.org/doi/10.1103/PhysRevB.54.16864}.

\end{thebibliography}

\end{document}